\documentclass[12pt]{article}
\usepackage{amsmath,amssymb,amsthm,textcomp,graphicx,subfigure}
\begin{document}
\baselineskip=0.8cm

\begin{center}
{\large Neptune migration model with one extra planet\\[0.5cm]
Lun-Wen Yeh$^{a, }$* and Hsiang-Kuang Chang$^{a, b}$\\[0.3cm]}
$^a$Department of Physics, National Tsing Hua University, Hsinchu, 
Taiwan 30013, R.O.C.\\
$^b$Institute of Astronomy, National Tsing Hua University, Hsinchu, 
Taiwan 30013, R.O.C.\\
*Corresponding Author E-mail address: d913316@oz.nthu.edu.tw
\end{center}
Pages: 42\\
Tables: 0\\
Figures: 12\\
\clearpage

\begin{flushleft}
{\bf Proposed Running Head:\\}
Neptune migration model with one extra planet\\[2.0cm]

{\bf Editorial correspondence to:\\}
Lun-Wen Yeh\\
Phone: +886920039692\\
E-mail address: d913316@oz.nthu.edu.tw
\end{flushleft}
\clearpage

\begin{flushleft}
{\large\bf ABSTRACT}\\
\end{flushleft}

We explore conventional Neptune migration model with one additional planet 
of mass at 0.1-2.0 $M_\oplus$. This planet inhabited in the 3:2 
mean motion resonance with Neptune during planet migration epoch, 
and then escaped from the Kuiper belt when Jovian planets parked 
near the present orbits. Adding this extra planet and assuming the 
primordial disk truncated at about 45 AU in the conventional Neptune migration 
model, it is able to explain the complex structure of the observed 
Kuiper belt better than the usual Neptune migration model did in several 
respects, which are the following. (1) High-inclination Plutinos with 
$i\simeq15${\textdegree}-35{\textdegree} are produced. (2) 
Generating the excitation of the classical Kuiper belt objects, which have 
moderate eccentricities and inclinations. (3) Producing the larger 
ratio of Neptune's 3:2 to 2:1 resonant particles, and the lower ratio 
of particles in the 3:2 resonance to those in the classical belt, which
may be more consistent with observations. (4) Finally, several Neptune's 
5:2 resonant particles are obtained. However, numerical experiments imply 
that this model is a low-probability event. 
In addition to the low probability, two features produced by this 
model may be inconsistent with the observations. They are
small number of low-inclination particles in the classical belt,
and the production of a remnant population with near-circular and
low-inclination orbit within $a\simeq50$-52 AU.
According to our present study, including one extra planet
in the conventional Neptune migration model as the scenario we explored 
here may be unsuitable because of the low probability, and the two drawbacks
mentioned above, although this model
can explain better several features which is hard to produce by the conventional 
Neptune migration model. The issues of low-probability event and the lack
of low-inclination KBOs in the classical belt are interesting and may be studied further 
under a more realistic consideration.

\begin{flushleft}
\end{flushleft}
\clearpage

\section*{\bf 1. Introduction}

\qquad
Kuiper belt objects (or Edgeworth-Kuiper belt objects) are the relics 
of primordial planetesimal disk beyond the orbit of Neptune in our 
solar system. More than one thousand Kuiper belt objects (KBOs) have 
been observed. Dynamical models for the origin and orbital evolution 
of the KBOs have been proposed by many authors (For a short review of 
various dynamical models, see e.g. Lykawka and Mukai, 2008; 
Gomes, 2009). One of the models is 
Neptune migration model (Malhora, 1995; Hahn and Malhotra, 2005), 
which invokes four migrating Jovian planets in a swarm of primordial 
planetesimals to explain the present structure of the Kuiper belt. 
Hahn and Malhotra's (2005) model is the most updated one and their 
results can explain many important observed features of the KBOs. However,
their model still have several weaknesses, which are (1) In 
accretion models (see e.g. Kenyon and Luu, 1999; Kenyon, 2002),  
KBOs were formed in a cluster of small bodies with initially nearly 
circular and coplanar orbits (eccentricity $e$ and inclination 
$i\lesssim10^{-3}$). Hahn and Malhotra's model assumes an initially 
dynamically hot disk (the mean value of $e$ and $i\simeq0.1$). However,
the stirring mechanism for these hot particles is unclear; (2) Few 
high-inclination KBOs were produced by their model. It cannot 
account for observations, especially for the KBOs at classical 
belt (the non-resonant KBOs with the semimajor axis $a$ in the range of 
$37\lesssim a\lesssim 50$ AU and the perihelion  $q\gtrsim37$-40 AU) 
and at Neptune's 3:2 mean motion resonance (MMR); and (3) At the end 
of their simulation there are plentiful KBOs at Neptune's 2:1 MMR, 
which may be inconsistent with observations. The de-bias number ratio 
between Neptune's 3:2 and 2:1 resonant KBOs inferred from 
the observation by Lykawka and Mukai's study (Lykawka and Mukai, 2007a) 
is 2.8, but that ratio in Hahn and Malhotra's simulation is about 
0.4. The simulated ratio of the resonant KBOs can be influenced by several 
processes which were not included in their model, for example, the 
stochastic effect during planet migration (Zhou et al., 2002), 
the different migration timescales for the artificial force that drives 
the migration of Jovian planets (Chiang and Jordan, 2002), and the gas 
drag during resonance capture (Jiang and Yeh, 2004; de La Fuente 
Marcos and de La Fuente Marcos, 2008) .

To improve the migration model in respect of the afore mentioned three 
weaknesses, we employ one additional planet with 0.1 to 2.0 
$M_{\oplus}$ in the Neptune migration model. As in the Neptune 
migration model, we start with the four Jovian planets embedded in a 
swarm of dynamically cold particles, which represent the KBOs. 
At the beginning of planet migration this extra planet is located 
at the 3:2 MMR with pre-migrated Neptune, and then due to the resonance 
capture of Neptune the extra planet migrates outward together with 
Neptune. Theoretical predictions and observational implications for the 
existence of the planets with masses of about several tenths to several 
Earth mass in the early outer solar system have been studied in the 
literature (e.g., Stern, 1991; Fern{\'a}ndez and Ip, 1996). Moreover, 
the gravitational perturbation of the extra planet with mass of a few 
tenths to two Earth mass in the early trans-Neptunian space has 
already been studied in Petit et al. (1999) and Gladman and Chan 
(2006) for the situation without the planet migration. Petit et al. 
(1999) studied the perturbation of this planet-size object in the 
inner region of the Kuiper belt with semimajor axis $a$ within 30-50 
AU region, while Gladman and Chan (2006) explored that in the outer 
region of the Kuiper belt. Generally speaking, the existence of this 
massive object excites the primordial disk and produces 
high-inclination KBOs. Furthermore, in the conventional Neptune 
migration model, usually there are many particles trapped in 
Neptune's 2:1 MMR during migration. By including one extra planet in 
the 3:2 MMR with Neptune during migration era, the capture rate 
of Neptune's 2:1 MMR may be reduced because of the close encounter 
of the 2:1 resonant KBOs with this extra planet, and because of the 
overlap between the 2:1 MMR of Neptune and the 4:3 MMR of this extra 
planet. With the aid of the extra planet, we explore the possibility 
to improve the afore mentioned weaknesses in the conventional model. 
In our model we assume that this extra planet leaves the Kuiper belt
in a very short timescale near the end of Neptune migration, and 
we artificially remove it. Otherwise it will deplete most particles 
and destroy the belt's structure. Our integration continues to about 
0.5 Gyr, when the distribution of all test particles is roughly 
stabilized.

Recently, Lykawka and Mukai (2008) also studied a putative planet in 
the Neptune migration model. Their extra planet with 0.3-0.7 
$M_{\oplus}$ was located at Neptune-crossing orbit and has 
$a\simeq60$-80 AU before Neptune's migration. The extra planet was 
responsible for the excitation of the Kuiper belt and disk truncation 
at 48 AU before migration era. During or after the end of planet 
migration it was captured by one distant MMR with Neptune and 
inhabited in a stable orbit at $a\simeq100$-175 AU with suitable 
$e$ and $i$. In this paper, we investigate another possible but different 
scenario from theirs to explain the Kuiper belt's structure.

In the following, we use ``conventional migration model" and 
``extra-planet model" to represent the conventional Neptune migration 
model and our model, respectively. We describe numerical procedures 
in Section 2. In Section 3 we examine the influence of resonances 
overlap between Neptune's and the extra planet's MMRs on the 
capture of resonant particles. In order to single out this effect, no 
migration is adopted. The main simulations of the extra-planet model 
are described in Section 4. We report the major results of the simulation 
in Section 5, in which comparisons between simulation results and 
observations are also presented. In the final part of Section 5, 
we explore the likelihood of the orbital evolution of the extra-planet
invoked in the extra-planet model. The simulation results imply that this
model is a low-probability event. We also discuss two issues in the 
model results, which are lack of low-inclination particles in the 
classical belt and the production of a remnant population
with near-circular and low-inclination 
at near $a\simeq50$-52 AU. Conclusions are described in Section 6.

\section*{\bf 2. Numerical procedures}

\qquad
We used the hybrid symplectic algorithm in the Mercury 6, which is an 
N-body integrator (Chambers, 1999), to perform our simulations. 
In the numerical integrations, 
only gravitational force is involved.
Other physical processes such as accretion or fragmentation are 
neglected. 
In Section 3 and 4, our solar system consists of the Sun, 
the four Jovian planets, one extra planet and thousands of test 
particles with negligible masses. 
The four Jovian planets and the extra planet interact with one 
another but are not perturbed by the test particles. The motion of
individual particle governed by the gravitational force of  
the four Jovian planets and the one extra planet but not by other test 
particles. The test particles are discarded when colliding with any 
massive objects or going beyond heliocentric distance of 1500 AU. In 
our main simulations described in Section 4, the four Jovian planets 
migrate smoothly under the influence of the artificial force which is 
the same as that used in Hahn and Malhotra's study (2005). This 
artificial force represents the gravity of all test particles. 
This simplification is supported by a self-consistent simulation of 
Hahn and Malhotra (1999). Initially the four Jovian planets are 
posited at a more compact orbits and then forced by artificial 
velocity kick  ${\Delta}v$ in time step ${\Delta}t$ with the form
\begin{equation}
{\Delta}v = \frac{1}{2}\frac{{\Delta}a}{a}\frac{{\Delta}t}{\tau}e^{-
\frac{t}{\tau}}v,
\end{equation}
which causes the planet's semimajor axis to vary as 
\begin{equation}
a(t) \simeq a_f-{\Delta}ae^{-\frac{t}{\tau}}, 
\end{equation}
where $a$ is the semimajor axis of the planet, $a_f$ is the final 
semimajor axis of the planet, ${\Delta}a$ is the planet's radial 
displacement, ${\tau}$ is migration timescale. All 
simulations in Section 3 and 4, a time step of 0.5 year is 
used\footnotemark[1], whose adequacy is discussed by Hahn and 
Malhotra (2005).

\footnotetext[1]
{In Section 5.3.2, we use 1-year time step to reduce computing 
time for the simulations containing massive particles.\\}

\section*{\bf 3. Numerical examination for resonances overlap}  

\qquad
In the extra-planet model the orbital configuration of Neptune, 
the extra planet and the Twotinos (the particles in the 2:1 MMR with Neptune)
forms a three-body mean motion resonance, which satisfies the relation 
of $n_N-3n_E+2n_T\simeq0$, where $n_N$, $n_E$ and $n_T$ are the 
mean motion of Neptune, the extra planet, and the Twotinos, 
respectively and $n_N:n_E:n_T=6:4:3$. The orbits of the three-body 
mean motion resonance are usually chaotic (Murray et al., 1998; 
Nesvorn{\'y} and Morbidelli, 1998a; Nesvorn{\'y} and Morbidelli, 
1998b). Therefore, one additional planet located at Neptune's 
3:2 MMR during planet migration may reduce Neptune's 2:1 
resonant particles due to the resonance overlap between the 
2:1 MMR of Neptune and the 4:3 
MMR of the extra planet. The main purpose of this section is to 
examine this resonance overlapping effect numerically. To simplify 
the situation no planet migration is employed here. All runs in this 
section include four Jovian planets, 1000 test particles in the 3:2 
MMR and 1000 test particles in the 2:1 MMR with Neptune. The present 
orbits are used for the initial orbital elements of the four Jovian 
planets (obtained from the JPL HORIZONS system). The initial particles 
are uniformly distributed with semimajor axis $a\in$ (38.9, 40.1) 
AU and (47.2, 48.4) AU for Neptune's 3:2 and 2:1 MMR
respectively, and have eccentricity $e$ 
and inclination $i$ following the Rayleigh distribution with mean 
value $\langle e\rangle=0.1$ and $\langle$sin$i\rangle=\langle e\rangle$/2. 
The argument of pericentre $\omega$, the 
longitude of the ascending node $\Omega$ and the mean anomaly 
$\lambda$ are randomly chosen between 0{\textdegree} and 
360{\textdegree}. Three runs among them include one extra planet with 
mass 0.1, 0.5 or 1.0 $M_{\oplus}$ at the beginning of the integration, and 
one run without one extra planet is for comparison. One body in 
Neptune's $j+k:j$ MMR is defined as the body having librating 
resonant angle $\phi$ around some fixed value and with amplitude 
$\Delta\phi$; $\phi\equiv(j+k)\lambda -j\lambda_{N}-k\varpi$, where j, k 
are integer, $\lambda$ and $\lambda_{N}$ are the mean longitude of the 
body and Neptune, respectively, and $\varpi$ is the body's longitude 
of perihelion. Initially the extra planet is set 
at Neptune's 3:2 MMR ($a=39.5$ AU) with zero $e$ and $i$; 
$\omega$, $\Omega$ and $\lambda$ are chosen 
such that this planet has resonant amplitude $\Delta\phi$ $\le$ 
110{\textdegree}, which provides a more stable state at Neptune's 
3:2 MMR (Levison and Stern, 1995; Nesvorny and Roig, 2000). The 
destruction of the structure of Neptune's 3:2 MMR due to the perturbation 
of the extra planet can be demonstrated by 3:2 resonant particles. It 
gives a constraint on the duration of the extra planet's stay in the 
3:2 MMR with Neptune. For each run the total integration time is 8.25 
Myr ($=5\times10^4$ $T_N$, where $T_N$ is the orbital period of 
the present Neptune and is 165 yr). This integration time is much longer than 
the typical libration periods of Neptune 3:2 and 2:1 MMRs, which 
are about 100-1000 $T_N$.

Fig. 1 shows time-averaged semimajor axis $\langle a\rangle$ 
and eccentricity $\langle e\rangle$ over a time interval of 0.825 
Myr for the test particles and the extra planet at the end of 
the integration. Figs. 1a-1d plot the one run without extra planet and 
the three runs including one extra planet with the mass of 0.1, 0.5 and 1.0 
$M_{\oplus}$ respectively. We define the resonant particles to be 
those having resonant amplitude $\Delta\phi \le 170${\textdegree}. 
This criterion avoids to identify most of non-resonant particles as 
the resonant ones due to insufficient time-sampling.\footnotemark[2] 
At the end of the simulation, there are 15, 35, 205 and 214 Neptune's 2:1 
resonant particles for the runs containing one extra planet of 1.0 
$M_{\oplus}$, 0.5 $M_{\oplus}$ and 0.1 $M_{\oplus}$ , and without 
extra planet, respectively. These simulation results indicate that the 
2:1 resonant particle number decreases as the mass of the extra planet 
increasing. This tendency is because the particles near Neptune's 
2:1 MMR had close-encounters with the extra planet, and because of 
the overlap between Neptune's 2:1 and the extra planet's 4:3 MMR. 
Both of the effects are intensified as the mass of the extra planet 
increases. Below the equal-perihelion line of 40.4 AU in the 
$\langle a\rangle$-$\langle e\rangle$ diagram (the curved dashed 
line in Fig. 1) Neptune's 2:1 resonant particles suffered almost no 
close-encounter with the extra planet, but the 2:1 resonant particle 
number was still depressed as the mass of the extra planet increases. 
This feature displays the resonance overlapping effect more. As for 
Neptune's 3:2 resonant particles in the runs containing one extra 
planet, all of them had close approach with this extra planet during 
the integration, thus the decrease of Neptune's 3:2 resonant particles 
is expected. When the integration time extended to several tens of Myr 
for the 1.0-$M_{\oplus}$ extra-planet run, there was almost no test 
particle at or near Neptune's 3:2 MMR due to the extra planet's strong 
perturbation. To clarify the MMRs overlap further we perform another 
run for the 1-$M_{\oplus}$ extra-planet case where the initial 
semimajor axis of the extra planet is shifted by 1 AU farther
beyond the exact 3:2 MMR of Neptune. This selection avoids the 
first-order MMRs of the extra planet overlapping with Neptune's 
2:1 MMR. The $\langle a\rangle$ and $\langle e\rangle$ at 
the end of this simulation are demonstrated in Fig. 1e. At the end of 
the simulation, there are 53 particles trapped in Neptune's 2:1 MMR. 
It is more than those in Fig. 1d, where Neptune's 2:1 MMR captures
merely 15 particles, although the extra planet is closer to  
Neptune's 2:1 MMR in Fig. 1e. Moreover, below the equal-perihelion 
line of 40.4 AU the 2:1 resonant particles are more abundant in 
Fig. 1e than those in Fig. 1d. Therefore we conclude that the overlap 
of the 2:1 MMR of Neptune and the 4:3 MMR of the extra planet can 
reduce the resonant particles of Neptune due to the long-term gravitational 
perturbation of this extra planet. Reducing the resonant KBOs at  
Neptune's 2:1 MMR can be achieved by setting one extra planet at 
Neptune's 3:2 MMR during Neptune's migration. It may therefore
improve the conventional migration model. In the next section we explore 
this picture with planet migration.

\footnotetext[2]
{To know the influence of the time-sampling on resonant particle 
identification, we constructed a toy model using a sinusoidal curve 
to model particle's resonance libration. In this paper we define the 
particle with resonant amplitude $\Delta\phi\le 170${\textdegree}
as the resonant particle. The toy model indicated that this selection 
recognizes $\lesssim1$\% non-resonant particles as the resonant ones.}

{\bf [Figure 1]}

\section*{\bf 4. Planet migration with one additional planet}

\subsection*{4.1. Short-term (82.5 Myr) simulations}

\qquad
In this section we describe simulations with the planet migration. 
The semimajor axes of the four Jovian planets vary as Eq. (2). The 
initial semimajor axis and the radial displacement \{$a_0$, ${\Delta a}$\} 
for Jupiter, Saturn, Uranus and Neptune are \{5.4 AU, -0.2 AU\}, 
\{8.7 AU, 0.8 AU\}, \{16.2 AU, 3.0 AU\} and \{23.2 AU, 7.0 AU\}, 
respectively. These values have been inferred and adopted in several 
papers (Malhotra, 1995; Hahn and Malhotra, 1999; Chiang and Jordan, 
2002; Lykawka and Mukai, 2008). All other orbital elements for the 
Jovian planets are the same with the present values (adopted from the 
JPL HORIZONS system). The migration timescale $\tau$ we adopted is 
10 Myr (Hahn and Malhotra, 1999). In addition to the four Jovian 
planets, one extra planet is set at 
Neptune's 3:2 MMR at the beginning of the planet migration. The eccentricity 
and inclination of this extra planet are arbitrarily chosen but are 
similar to those of the four Jovian planets. Other orbital elements 
$\omega$, $\Omega$ and $\lambda$ are chosen such that the extra planet 
has smaller resonant amplitude $\Delta\phi$, which is usually less 
than 110{\textdegree} during our integrations. We also include 4000 
test particles initially distributed from $a=24.0$ AU to 50.0 AU with 
the surface number density varied as $a^{-2}$. The $e$ and $i$ of particles 
follow the Rayleigh distribution with mean value $\langle e\rangle=0.001$ 
and $\langle$sin$i\rangle=\langle e\rangle$/2. The $\omega$, 
$\Omega$ and $\lambda$ of the particles are randomly chosen within 0 
to 360{\textdegree}. To clarify that our final simulation results and 
conclusions do not depend on the particular choice of initial 
particle set and extra planet, we select five particle-set a, 
b, c, d and e which combine with five different extra-planet A, B, 
C, D and E, respectively. The initial semimajor axis, eccentricity and 
inclination in the form of ($a$, $e$, $i$) for the extra-planet A, B, C
,D and E are (31.0 AU, 0.019, 3.25\textdegree), 
(30.6 AU, 0.071, 1.50\textdegree), (30.6 AU, 0.029, 1.96\textdegree),
(30.9 AU, 0.016, 1.94\textdegree) and (30.9 AU, 0.029, 2.42\textdegree),
respectively. For each combination of one particle set and one extra 
planet, we perform three runs for the extra planet with mass of 0.1 
$M_{\oplus}$, 0.5$M_{\oplus}$, 1.0 $M_{\oplus}$. One run without extra 
planet is also executed. Thus, we totally execute twenty runs and the 
integration time for each run is 82.5 Myr ($=5{\times}10^5$ $T_N$), 
which is $\simeq8\tau$.\footnotemark[3] For all the simulations 
described in Section 4, the $\langle a\rangle$, 
$\langle e\rangle$ and $\Delta\phi$ are calculated over a time 
interval of 8.25 Myr ($=5{\times}10^4$ $T_N$).

Fig. 2 plots an example for the semimajor axes of the four Jovian 
planets and one extra planet varying with time in smooth planet 
migration, where the extra-planet A has the mass of 1.0 $M_\oplus$. 
The four giant planets were driven by the artificial force while the extra 
planet was trapped by the 3:2 MMR with Neptune and moved outward. The 
resonant particle numbers varying with time for the 3:2 and 2:1 MMRs with 
Neptune in the runs containing the particle-set a are shown in Fig. 3. 
In this figure we plots the three runs with the extra planets of different 
masses and the one run without extra planet. Roughly speaking, during 
the significant migration era of about 0-20 Myr resonant particle 
numbers increased due to resonance capture mechanism. As Neptune 
parked near 30 AU, many unstable resonant particles left the MMRs
gradually after about 20 Myr. Furthermore, there are several noteworthy 
features in this figure, (1) in the no extra-planet run, the time for 
the resonant particle numbers reaching the maximum value is late for the 
2:1 MMR compared to that of the 3:2 MMR. This phenomenon has been 
mentioned in the study of Zhou et al. (2002); and (2) in Fig. 3, the humps 
of the 2:1 and 3:2 resonant particles are depressed as one extra planet 
contained in the simulations. For the runs involving one extra planet 
with the masses of 0.5 $M_\oplus$ and 1.0 $M_\oplus$, the particle 
numbers trapped in the 2:1 MMR are always less than ten. Comparison 
with conventional migration model result (the run without one extra 
planet), the maximum resonant particle numbers of the 3:2 MMR reduce 
to 5.0\%, 5.6\% and 42\% for the runs containing one extra planet 
of 1.0, 0.5 and 0.1 $M_\oplus$, respectively. Similar quantities for 
the 2:1 MMR are 0.77\%, 0.66\% and 38\%. With the existence of the 
extra planet fewer particles were captured in the 2:1 MMR with Neptune 
because the MMRs overlapped between the 2:1 MMR with Neptune and the 4:3 
MMR with the extra planet, and Neptune's 2:1 resonant particles had
close-encounters with the extra planet. The decrease for the capture 
rate of Neptune's 2:1 MMR is more than that of Neptune's 3:2 
MMR as the mass of the extra planet increases. Other sixteen runs with 
different initial extra planets and particle sets gave similar features. 
Hence, We conclude that in the conventional migration model the ratio 
of Neptune's 2:1 to 3:2 resonant particles can be decreased due 
to the presence of one additional planet, which may improve the 
conventional migration model. This extra planet must escape during or 
near the end of the planet migration. Otherwise the structure of Neptune's 
3:2 MMR would be destroyed as a result of extra planet's perturbation. 
We assume that this extra planet leaves the Kuiper belt in a very 
short timescale. We remove it by hand and then continue the 
integration to about 0.5 Gyr in the next subsection.

\footnotetext[3]
{All the extra planets in these 82.5 Myr short-term runs were captured in 
the migrating Neptune's 3:2 MMR, except the run contains the extra-planet 
B with the mass of 0.1 $M_\oplus$. In this run the extra planet 
escaped from the 3:2 MMR of Neptune at about 23 Myr.}

{\bf [Figure 2, Figure 3]}

\subsection*{4.2. Long-term (495 Myr) simulations}

\qquad
From the previous twenty short-term runs with about eight times migration 
timescale, we have understood the time variation of the resonant particles. 
Among the twenty runs we reproduce fifteen runs containing one extra 
planets, and then remove the extra planet by hand in each run at 
certain times during or near the end of the Neptune migration. After 
the removal of the extra planet, the integrations are continued to 495 
Myr (= $3{\times}10^6$ $T_N$). Based on Fig. 3, the particular times 
that we adopt to discard one extra planet are 8.25, 24.75, 33.0 and 
49.5 Myr (which are $5{\times}10^4$, $1.5{\times}10^5$, 
$2{\times}10^5$, $3{\times}10^5$ $T_N$, respectively). In addition, 
we also extend the remaining five runs without extra planet to 
495 Myr for comparison. We will discuss the results for the 
conventional migration and the extra-planet model in Section 5.

\section*{\bf 5. Results and discussion}
\subsection*{5.1. Results of the 495 Myr integration and truncated disk}

\qquad
Taking the run containing the particle-set a as an example, we 
removed the 1-$M_\oplus$ extra planet at 33 Myr, which is roughly near the 
end of the migration era, and expect that the ratio of Neptune's 
2:1 to 3:2 resonant particle numbers will be reduced in the end of 
the 495 Myr simulation comparing to the conventional model results. The 
right panel of Fig. 4 shows the numbers of the resonant particles varying 
with time for this long-term integration. The left panel of Fig. 4 
is just an enlarged of part of Fig. 3 for the run containing 
the 1.0-$M_\oplus$ extra planet trapped in the 3:2 MMR with Neptune 
until 82.5 Myr integration. After removal of this additional planet, 
Neptune's 2:1 resonant particle numbers increased unexpectedly 
in several tens Myr and were in excess of the 3:2 resonant particle 
numbers eventually at the end of 495 Myr (the right panel of Fig. 4). 
The abrupt increase of the particle number in the 2:1 MMR with 
Neptune is because that at the time we removed the extra 
planet the test particles around Neptune's 2:1 MMR were abundant. 
As a consequence, those particles were captured by the 2:1 MMR 
immediately. The top panels of Fig. 5 demonstrates the orbital 
distribution of the test particles at the end of 495 Myr for this 
simulation, and the bottom panels shows those for the run without 
extra planet. The observed KBOs with multi-opposition are also 
plotted in Fig. 5. In the extra-planet model results, the excitation 
of initially cold test particles due to the perturbation of 
1-$M_\oplus$ extra planet is apparent in $a$-$e$ and $a$-$i$ 
diagrams. But there are still plentiful resonant particles in 
Neptune's 2:1 MMR.

{\bf [Figure 4, Figure 5]}

Inspecting the results more carefully, we find other discrepancies 
from the results in the extra-planet model and those in the traditional 
migration model. On the $a$-$e$ distribution of Fig. 5 we label the 
test particles according to their initial semimajor axes (Fig. 6). 
The colors in the $a$-$e$ diagram of the conventional migration model 
(the right panel of Fig. 6) show that the 2:1 MMR with Neptune 
contains the particles with initial semimajor axes among 35-50 AU. 
However, in the $a$-$e$ diagram of the extra-planet model 
(the left panel of Fig. 6) most particles in the 2:1 MMR have initial 
semimajor axes $a$ within 45-50 AU. As predicted in resonance 
capture theory, the eccentricities of test particles captured in 2:1 
MMR get larger as the semimajor axes increase during Neptune's migration. 
Therefore, in the right panel the 2:1 resonant particles with high 
eccentricities originated from interior region and those with low
eccentricities originated from exterior region. In the left panel of 
Fig. 6, this feature is not clear as one extra planet with 
1-$M_\oplus$ mass is involved in the model. Most of the 2:1 
resonant particles originated from the primordial disk of 45-50 AU.  
In the 2:1 MMR of the extra-planet model, the absence of 
particles with the initial semimajor axis $a$ smaller than 45 AU 
is due to following causes. First, the resonances overlapped at the 
2:1 MMR with Neptune and the 4:3 MMR with the extra planet during the 
migration era, which reduced the capture rate of the MMR as we 
demonstrated in Section 3. Second, some parts of the Neptune's 2:1 
resonant particles were destroyed by the close-encounters between the 
extra planet and the particles. Fig. 6 indicates a possible way to reduce 
the particles in Neptune's 2:1 MMR under the scenario of the 
extra-planet model. The observed KBOs indicates an 
edge with semimajor axis $a$ between 45-50 AU (Trujillo and Brown, 
2001a). In the extra-planet model if we truncate the initial particle disk 
at 45 AU, the 2:1 resonant particle number becomes small naturally 
(Fig. 7). During the migration epoch, the 2:1 MMR with Neptune shifted 
through the region of $a\simeq36$ to 47 AU. In the conventional 
migration model, many particles within the above region were trapped 
by the 2:1 MMR and finally there are plenty of 2:1 resonant 
populations at the simulation end. However, in the extra-planet model 
with the primordial disk truncated at 45 AU, the presence of the extra 
planet prevented most capture events in the 2:1 MMR when the 2:1 MMR 
passed through $a\simeq36$-47 AU in the first several tens Myr 
of Neptune's migration. With the assumption of the primordial disk's 
edge, we can compare the results of the extra-planet model with 
the observations. It provides some constraints on this model.

{\bf [Figure 6, Figure 7]}

\subsection*{5.2. Comparison with observed KBOs}

\subsubsection*{5.2.1. $a$-$e$ and $a$-$i$ distributions}

\qquad
Fig. 7 shows that the simulation results at the end of 495 Myr for 
the runs containing the particle-set a in the extra-planet model 
and the conventional migration model with the initially truncated disk 
at 45 AU. The extra planet was removed at 33 Myr in the extra-planet 
model run. At about 0.5 Gyr the region beyond Neptune is roughly 
stabilized. Hence, we compare simulating particles distribution with 
those of the present KBOs. Observed KBOs with multi-opposition are plotted 
in Fig. 7. At first glance compared with the results of the 
traditional migration model (the bottom panels of Fig. 7), the particle 
orbital distributions of the extra-planet model (the top panels of Fig. 7) 
are more consistent with observations in several respects, (1) the 
classical belt has notable 
excitation in $a$-$e$ and $a$-$i$ spaces due to the perturbation 
of the extra planet during the migration era; 
(2) the final 2:1 resonant KBOs number is few due to the MMRs 
overlap, the closes-encounters during the migration epoch and the assumption 
of the primordial disk edge; and (3) the inclinations of the Plutinos (the 
KBOs at the 3:2 MMR with Neptune) distribute in a very wide range 
with maximum value $\simeq30${\textdegree}. These three features 
can be seen in the other runs with extra planet having the same mass 
and the same escaping time. One main discrepancy between the 
extra-planet model results and the observations is that there are only 
few classical KBOs with inclination $i\lesssim5${\textdegree} 
(the top-right panel of Fig. 7). Most telescopes observed the KBOs near
ecliptic. The probability of the detection of the KBOs
at the ecliptic is 
roughly proportional to 1/sin$i$. Therefore, low-$i$ KBOs spend more time near 
the ecliptic and are more easily observed by telescopes than high-$i$ KBOs. 
The comparison of KBO inclination distribution between 
simulation's and the observation's should consider this telescopic 
selection effect. Fig. 8 shows the ecliptic inclination distribution 
for the simulated particles and the observed KBOs. The ecliptic inclination 
distribution is the inclination distribution of particles with 
latitude $\beta$ near the ecliptic, which can mitigate above selection 
effect (Brown, 2001; Hahn and Malhotra, 2005). We 
choose particles with $\beta\le3.0${\textdegree} and perihelion 
$q\le45.0$ AU.\footnotemark[4] As previous understanding of the 
conventional migration model, it produces deficient high-inclination 
KBOs with inclination $i\gtrsim5${\textdegree} (the right panels 
of Fig. 8). In addition, the simulated particles with inclination 
$i\lesssim5${\textdegree} are too many. This is mainly due to the 
initially dynamically cold particles. As for the ecliptic inclination 
distribution of the extra-planet model, the simulated particles can 
cover most of observed high-inclination KBOs with $i\gtrsim5${\textdegree}. 
Nevertheless, the simulated particles with $i$
within 10{\textdegree}-20{\textdegree} are a little more than the 
observed KBOs, and  the low-inclination KBOs with $i\lesssim5${\textdegree} 
are few in the extra-planet model's results.

\footnotetext[4]
{In the simulation end few particles orbited near the ecliptic in the  
extra-planet model results. In order to enhance the statistic meaning 
in the ecliptic inclination distribution, we adopted the particles 
with $\beta\le3.0${\textdegree} to plot the ecliptic inclination 
distribution. The choice of the perihelion 45 AU roughly equals to 
observation limit.}

{\bf [Figure 8]}

Under the picture of the extra-planet model with one escaping planet 
from Neptune's 3:2 MMR and an initial disk truncated at $a\simeq45$  
AU, we can roughly constrain the mass and the removal time of this 
extra planet from the observational KBOs' orbital distribution. 
Among all our runs, the model employing 1-$M_\oplus$ extra planet 
with the removal time near the end of the Neptune migration (33.0 Myr) 
gives the most consistent final particle distribution, because of 
following several reasons. (1) For the simulations with one extra 
planet of the same mass (1 $M_\oplus$), the extra planet inhabiting 
in the 3:2 MMR too long (0-49.0 Myr) consumed most of the 3:2 
resonant particles and produced particles with too high inclination 
for the ecliptic inclination distribution, while it inhabiting in the 
3:2 MMR too short (0-8.25 Myr) had insufficient perturbation of the 
Kuiper belt. (2) As for the simulation with the same removal time for 
the extra planet but with different masses, the excitation of the 
Kuiper belt was not adequate for the model with lower mass extra 
planet (0.1 and 0.5 $M_\oplus$). We perform another run with 
2.0-$M_\oplus$ extra planet to 82.5 Myr. At 33.0 Myr there were
almost no classical KBOs with $i\le6${\textdegree}, which cannot 
account for the observations. (3) Increasing the extra planet's mass 
and postponing its escaping time simultaneously overly excited the 
belt, while decreasing the extra planet's mass and removing it earlier 
together had inadequate perturbation on the belt.
(4) The model including the extra planet having larger mass 
(2.0 $M_\oplus$) with shorter inhabited time in the 3:2 MMR avoided
depleting all low-inclination classical KBOs, however its 2:1 MMR 
with Neptune captured too many particles during the migration era after 
the discard of the extra planet. (5) The model with a lower mass 
extra planet (0.1 and 0.5 $M_\oplus$) cannot account for 
high-inclination Neptune's 3:2 KBOs, even if the integration time 
was extended to 82.5 Myr ($\simeq8\tau$). At this time,  
Neptune's 3:2 resonant particles became too few. Hence, it is 
inconsistent with observations.

In the previous discussion we only posited the extra planet in  
Neptune's 3:2 MMR at the beginning of the Neptune migration. We 
also attempt to put this extra planet in the 2:1 MMR with Neptune, 
which is another strong MMR of Neptune in the Kuiper belt. We do
three more runs, where this extra planet is at Neptune's 2:1 MMR 
when the Neptune's migration starts. The same as previous runs we 
remove the extra planet at or near the end of the migration and then 
integrate to 495 Myr. The main drawback of this configuration is 
that at the simulation end the dearth of the Plutinos of inclination 
$\ge27${\textdegree} is obvious in all the three runs. It is 
inconsistent with the observations. Therefore, we think that setting 
one extra planet at Neptune's 2:1 MMR is unlikely.

Base on the above discussion, according to the $a$-$e$, $a$-$i$ diagrams, 
ecliptic inclination distribution and Neptune's 3:2 to 2:1 
resonant particles' ratio, the mass and the removal time of the extra
planet are about 1.0 $M_\oplus$ and near the end of the migration 
($\sim$ 33 Myr). In our simulations, we merely performed several 
sparse points in a large parameter space for the mass and the removal 
time of the extra planet. Therefore, these two quantities can be only 
crudely determined. In the following subsections we will compare 
more simulation results of the runs adopting the extra planet with 
1.0-$M_\oplus$ mass and 33 Myr removal time with the observation KBOs.

\subsubsection*{5.2.2. Plutinos and other resonant KBOs}

\qquad
In Lykawka and Mukai's (2007a) identification of observed KBOs, 
there are 100 Plutinos among 622 KBOs. The number of the Plutinos 
provides a large fraction among total KBOs' population. Therefore, 
creating Plutinos with similar orbital distributions and physical 
properties to observations is an important task to the theoretical 
models. Fig. 9 demonstrates simulated time-averaged eccentricities 
$\langle e\rangle$, time-averaged inclinations 
$\langle i\rangle$ and resonant angles $\Delta\phi$ for the Plutinos. 
Those quantities for observed Plutinos are also plotted in this figure 
(from Lykawka and Mukai, 2007a). The main discrepancy between the simulated 
results of the conventional migration model and those of the extra-planet 
model is that the extra-planet model produced many high-inclination 
Plutinos with inclination between 15{\textdegree}-35{\textdegree}, 
which usually cannot perform by the conventional migration model 
(the top two panels of Fig. 9). These high-inclination Plutinos were 
generated due to the perturbation of extra planet and gives a more 
consistent $\langle e\rangle$-$\langle i\rangle$ distributions 
with the observations. In the bottom two panels of Fig. 9, some parts of 
the simulated Plutinos of the extra-planet model possess 
$\Delta\phi\ge130${\textdegree} that ought to be insufficient integration time 
in our simulation (Nesvorny and Roig, 2000). Generally, the orbital 
properties of the Plutinos generated from the extra-planet model are 
agreeable to the observations.

As for the ratio of Neptune's 3:2 to 2:1 resonant particles, 
the average value of this ratio provided by the five runs of the 
extra-planet model is 2.0. The same quantity provided by 
the five runs of the conventional migration model is 0.44.
Therefore, the ratio calculating from the extra-planet model is 
about five times larger than that of the conventional migration model,
and is more consistent with the observations ($\sim$ 2.8). Although in our 
simulations we assumed that migration process is artificially smooth, 
more realistic migration may not change this conclusion.

Another important ratio is the Plutino number to the number of the 
classical KBO. In the simulation end, we consider the particles having 
perihelion $q\ge37$ AU and $a\le50$ AU as the classical KBOs, 
except those belong to the 3:2, 5:3, 7:4 and 2:1 MMRs. The average ratio 
of the Plutino to the classical KBO numbers is 1.5 for the five 
runs of the conventional migration model, while the same ratio for the 
extra-planet model is 0.098, which is closer to a de-bias 
observational ratio that is about 0.04 (Trujillo and Jewitt, 2001b).

Beyond $a=50$ AU there are still abundant observed KBOs inhabiting 
in higher order MMRs with Neptune, e.g. the 7:3, 5:2, 3:1 and 4:1 MMRs. From 
Lykawka and Mukai's (2007a) study, among 622 KBOs there are 12 KBOs 
located in the 5:2 MMR with Neptune, which is the major resonant 
population outside $a=50$ AU. These distant resonant populations 
can be captured more easily during the Neptune migration if they were 
dynamically hot before resonance capture (Chiang et al., 2003; Hahn 
and Malhotra, 2005; Lykawka and Mukai, 2007b). In the extra-planet 
model, the perturbation of the extra planet supplies the excitation 
for particles in eccentricities and inclinations prior the resonance 
capture. Fig. 10 demonstrates that in the five runs of 
the extra-planet model, four runs have several Neptune's 5:2 resonant 
particles (left two panels), while only one among the five runs of the 
conventional migration model produces 5:2 resonant particles (right 
two panels). Therefore, more particles were trapped at Neptune's 
5:2 MMR in the extra-planet 
model than those in the conventional migration model. Generally, the 
behaviors of simulated 5:2 resonant particles of the extra-planet 
model are similar with the observations, except the inclination 
distribution of the run containing the particle-set a (the crosses in the top-left 
panel of Fig. 10). For this run the 5:2 resonant particles' 
inclinations are within 20{\textdegree}-35{\textdegree} 
which is higher than the observations. One possible explanation is that 
the observations prefer low inclination objects as we mentioned 
previously. For circular orbit objects, the detection probability 
near the ecliptic is about three times higher for the objects with 
$i=10${\textdegree} than for the objects with $i=30${\textdegree}. 
One run of the conventional migration model also provided 5:2 resonant 
particles but their resonant amplitudes are all larger than 
120{\textdegree}. None of the runs of the conventional migration model  
produced 5:2 resonant particle with small resonant amplitude, which 
has more stable orbit.

{\bf [Figure 9, Figure 10]}

\subsubsection*{5.2.3. Extended scattering KBOs}

\qquad
Usually extended scattering KBOs have $a\gtrsim50.0$ AU and 
$q\gtrsim40.0$ AU (definition varies slightly in literature). These 
objects do not suffer close-encounters of the giant planets and form one 
class of KBOs. From the Minor Planet Center database (July, 10, 2007), 
among 717 observed KBOs ($a\ge30$ AU) with multi-opposition
there are 7 extended scattering KBOs, e.g. 2004 $XR_{190}$ 
($a=57$ AU, $q=51$ AU), 2000 $CR_{105}$ ($a=218$ AU, $q=44$ AU) 
and Sedna ($a=487$ AU, $q=76$ AU). A biased abundance of this 
class KBOs is $7/717\simeq1$\%. The intrinsic ratio of the extended 
scattering KBOs should be larger than this value because of the larger 
distant of these objects. From the study of Gladman and Chan (2006), 
emplacing an Earth mass plant in the Kuiper belt promotes the 
production of the extended scattering KBOs with 50 AU 
$\lesssim a\lesssim500$ AU within several hundred Myr due to the secular 
perturbation of this extra planet. In the extra-planet model results, 
merely several particles have orbits like 2004 $XR_{190}$ with
50 AU $\lesssim a\lesssim80$ AU (Fig. 13). The absence of 
objects with orbits like 2000 $CR_{105}$ or Sedna is mainly due to 
that we assumed our extra planet escaping quickly near the end of 
the Neptune migration. Hence, the extra planet had insufficient time to 
influence the Kuiper belt. That the extra planet stay in the 
Kuiper belt more than several tens Myr would destroy the stability of 
Neptune's main mean motion resonance, for example, the 3:2, 5:3 and 
2:1 MMRs. Therefore, we may not invoke the picture in Gladman and 
Chan (2006) to produce 2000 $CR_{105}$-like or Sedna-like extended 
scattering KBOs in our model. Other possible mechanisms may be 
responsible for the construction of the extended scattering KBOs, 
including stellar passage (Ida et al., 2000; Kenyon and Bromley, 
2004) or a distant undiscovered planet (Lykawka and Mukai, 2008).

\subsubsection*{5.2.4. Outer edge of Kuiper belt}

\qquad
In the extra-planet model we assume the primordial planetesimal disk 
with an outer edge near 45 AU in order to decrease Neptune's 2:1 
resonant particles in our scenario. Another motivation is to have a 
more consistent $a$-$e$ distribution of the KBOs between model results 
and observations. The observed $a$-$e$ distribution implies an edge 
at $\simeq45$ AU. In the migration era of our model, the extra 
planet's semimajor axis and eccentricity increased simultaneously 
through migration. Although this extra planet always inhabited at the 
3:2 MMR, some initially cold particles within 40 AU 
$\lesssim a\lesssim45$ AU were still disturbed by high-eccentric migrating extra 
planet. This causes that some particles moved outward beyond 
$a\simeq45$ AU during extra planet's migration. The extra planet had
the final aphelion $\simeq$ 48 AU at the migration end and cleaned most 
of particles with near circular orbits within 45 AU 
$\lesssim a\lesssim50$ AU. Therefore, there is a group of nearly circular 
particles which still assemble within $a\simeq50$-52 AU (see 
the right panels of Fig. 11). 
We will discuss whether this remnant creating by the extra-planet 
model could be observed in Section 5.3.4.

{\bf [Figure 11]}

\subsection*{5.3. Discussion}

\subsubsection*{5.3.1. Assumption of truncation disk}

\qquad
In the extra-planet model, we assume the primordial disk having an 
edge at about 45 AU. Several processes may produce this edge prior 
to planet migration (see Gomes et al., 2004 and references therein),
for example, a passing star or nearby stars photoevaporated.

\subsubsection*{5.3.2. Removal of extra planet}

\qquad
In our simulation the extra planet was discarded artificially. We 
postulate that it leaves the Kuiper belt within a short timescale 
less than several Myr and does not influence the KBOs anymore. We 
execute 500 runs to investigate the possibility of this event. In 
each run we merely consider the Sun, four Jovian planets, and 
one 1-$M_\oplus$ extra planet at pre-migrated Neptune's 3:2 MMR 
with zero $e$, $i$ and different $a$ within 0.6 AU of exact resonance 
position. Other three orbital angles are randomly chosen between 
0{\textdegree} and 360{\textdegree}. Test particles are not included. 
Other orbital configurations for the Jovian planets and integration time 
are the same with the runs in Section 4.1. After 82.5 Myr 
integration, there are 486 survived extra planets, which are all have 
perihelion $q\le50$ AU and will still influence KBOs. Within 
82.5 Myr, only 13 extra planets escaped from the solar 
system\footnotemark[5] due to the close approach of 
one or multi Jovian planets, and the remaining one collided with 
Jupiter. The resonant amplitudes $\Delta\phi$ of these 14 objects 
are all larger than 90{\textdegree} in the early stage of migration.  
Therefore, in the extra-planet model if we invoke an extra planet 
leaving the Kuiper belt quickly near the end of the Neptune migration, 
the probability is less than $14/500\simeq0.03$.

The assumption of particles with negligible mass in the our simulation
loses gravitational effect on the extra planet. A 1-$M_\oplus$ mass 
extra planet embedded in a primordial planetesimal disk with mass 
of several tens $M_\oplus$ would suffer the dynamical friction, 
which circularizes the orbits of the extra planet and may reduce 
the chance to scatter the extra planet by Jovian planets. 
Furthermore, considering massive particles renders the stochastic 
migration, which decreases the capture probability for the extra 
planet during the migration (Zhou et al., 2002; Chiang et al., 2007).   
To explore these phenomena, we perform several runs using massive
particles. We do six runs with 4000 equal-mass particles, which 
distribute with 10 AU $\le a\le45$ AU, and small $e$ and $i$ 
as in Section 4.1. The surface mass density $\sigma$ = 
0.14($a$/40 AU)$^{-2}$ g cm$^{-2}$, which provides mass of $\simeq80$ 
$M_\oplus$ for initial disk. These particles interact with planets but 
not themselves. The initial orbits for the four Jovian planets are 
the same as those in Section 4.1, but no artificial
force is adopted here. One extra planet with 1-$M_\oplus$ mass having 
zero $e$, $i$ and different $a$ in each run is set at the 3:2 MMR with 
pre-migrated Neptune. In addition, we perform another same six runs 
but with 16000 equal-mass particles, which gives higher resolution. 
The total integration time for each run is 82.5 Myr. In all twelve 
runs, none of the extra planets was trapped by Neptune's 3:2 MMR 
during the main migration epoch of $t\lesssim$ 20 Myr. Most of the 
extra planets were captured when they migrated near the end of disk. 
The left and middle panels of Fig. 12 show two typical examples for the 
runs containing 4000 and 16000 massive particles, 
respectively\footnotemark[6]. We 
think that the extra planet was hardly captured in the 3:2 
MMR in the early epoch of the migration is because of large difference 
between the migration speed of the extra planet and that of 
the 3:2 MMR. Hence, the extra planet merely migrated across the 3:2 MMR.
When the extra planet was near disk edge, both the migration speed of 
the 3:2 MMR and the extra planet are slow, and then the extra planet 
was captured more easily. 
Furthermore, we find that the capture occurring after 20 Myr becomes 
easier when each particle possesses smaller mass. We believe that
this is due to the enhancement of capture ability as the Neptune migration
is more smooth (Chiang et al., 2007). In our higher resolution 
runs with the 16000 particles, 
each particle has mass of $5\times10^{-3}$ $M_\oplus$, 
which is still larger than reality by several orders 
(mass $\simeq10^{-6}$ 
$M_\oplus$ for 100 km size object). Instead of increasing 
particle number, which requires huge computing time, we reduce the 
total 
mass of primordial disk to investigate a less noisy migration for 
planets. 
We perform same six runs as previous with 16000 particles for each run, 
but initial disk mass has only 
15 $M_\oplus$. In all the six runs, at the beginning the extra planet was captured 
by Neptune's 3:2 MMR until about 10 Myr; after that, the 
extra planet left MMR and migrated outward to about 40 AU. The right panels of Fig. 12 
show one run with low-mass planetesimal disk.
We think whether the extra planet can be captured or not depends on
the smoothness of the Neptune migration, the dynamical friction on the  
extra planet, and the migration speed of the 3:2 MMR and that of the extra 
planet. Our simulations here merely imply that the capture of the extra planet 
by the 3:2 MMR during migration may be possible in a more realistic simulation. 
Concerning for the escape of the extra planet in the extra-planet model,
the extra planet in the massive-disk simulation indeed has larger 
eccentricity during capture state, but the value of $e$ still 
$\lesssim0.1$, which is hard for its scattering by the Jovian planets.
It is unclear that in a more realistic case how large the eccentricity can
reach during capture event.

In conclusion, more study is needed to better understand the interesting 
orbital evolution of the extra planet invoked in the extra-planet model 
in the future. In the present work, we just consider it as a low-probability 
event.

\footnotetext[5]
{The extra planet was removed from the integration as its heliocentric 
distance is large than 1500 AU.}

\footnotetext[6]
{The main purpose of the massive particle runs is to explore the 
extra-planet's dynamics in a massive disk. In Fig. 12, Neptune's 
final position does not locate near 30 AU. Parking Neptune at
suitable position depends on the initial orbital configuration of the 
four Jovian planets and the behavior of primordial disk. We think our 
main conclusions in the difference between the convention migration 
and the extra-planet model results would be retained under reasonable
initial planets' orbits and particle disk.}

{\bf [Figure 12]}

\subsubsection*{5.3.3. Issue of low-inclination classical KBOs}

\qquad
Although the extra-planet model can generate several 
observational features which  
are hard produced in the conventional migration model, there is still 
one main discrepancy between the extra-planet model results and 
the observations. The 
extra-planet model produces too few low-inclination KBOs with 
$i\lesssim5${\textdegree} in the classical belt, even if the ecliptic 
inclination distributions are used for comparison. In this model, we used 
massless particles to represent the KBOs in our simulations. At 
the time of several 
tens Myr that the extra planet was removed, $\sim60$\% of initial 
particles remains in the trans-Neptunian region. If the initial 
primordial disk has mass of 30 $M_\oplus$ (within 24-45 AU), these remaining 
particles have mass of $\sim$ 18 $M_\oplus$. From accretion theory, only 
$\sim2$-5\% of these particles/planetesimals have size $\sim$ 100 km, 
while the major disk mass was occupied by the particles/planetesimals
 having size with $\sim$ 0.1-10 km, which will then collide to 
dust grains and are removed from the Kuiper belt in Gyr timescale
(Kenyon et al., 2008). In this period these 100 km size objects 
suffer dynamical friction from small bodies. The cooling timescale 
of the dynamical friction for the big bodies is 
\begin{eqnarray}
t_{dy}\sim
i^{4}\frac{M_{\odot}^{1.5}}{M\sigma(Ga)^{0.5}f}\simeq
5\times10^{4}(\frac{0.6}{f})
(\frac{i}{5^{\circ}})^{4}(\frac{10^{-6}M_\oplus}{M})
(\frac{a}{40AU})^{1.5}\,Gyr,
\end{eqnarray}
where $i$ and $a$ are inclination and semimajor axis of big body,
respectively;
$\sigma$ = 0.14($a$/40 AU)$^{-2}$ g cm$^{-2}$ is the 
surface mass density of initial disk;
$M$ is the mass of big body, which is $\simeq$ $10^{-6}M_{\oplus}$ for 
the body of 100-km size; 
$f$ is the ratio of remaining disk as we remove the extra planet;
$G$ is the gravitational constant. 
This cooling timescale of $5\times10^4$ Gyr is much larger than 4.5 Gyr, the age of our 
solar system.
Therefore, the influence of the dynamical friction may not compensate for the shortage
of the low-inclination particles. Whether physical collision between the bodies of 
100-km size and numerous bodies of smaller size with total mass of about tens Earth mass 
alleviates the deficiency of the low-inclination part may be 
addressed in the future works.

\subsubsection*{5.3.4. Remnant population near $a\simeq50-52$ AU}

In the extra-planet model results of Fig. 11, the remnant populations
near $a\simeq50$-52 AU have near-circular and low-inclination
orbits with $e\lesssim$ 0.05 and $i\lesssim5$\textdegree, respectively.
To estimate how many objects in the remnant may be observed so 
far, we choose another cold classical population with $a$ within 42-45 AU, 
$i\le5$\textdegree and 
$q\ge40$ AU in the results of the extra-planet model. 
This cold classical population has similar small $e$ and $i$ with the above remnant 
population, but smaller heliocentric distance. In the our five runs of the extra-planet
model, the average number ratio of the remnant population to this cold classical
population is 0.93. With this ratio, and assuming these two populations have
equal size objects and average semimajor axes $43.5$ AU and $51.0$ AU,
we can roughly estimate the observational 
number of objects for the remnant population to be     
\begin{equation}
\frac{N_r}{N_c}\times\left(\frac{51.0}{43.5}\right)^4=0.93,
\end{equation}
where $N_r$ and $N_c$ are the observational number of objects for the remnant population and 
the cold classical KBOs under above criterion; $N_c=174$ from the Minor Planet Center database 
July, 10, 2007 for the KBOs with multi-opposition;
the second term in left hand side is the correction factor due to the observational bias
of flux. The $N_r$ provided by this estimate is 86, which is inconsistent with the observations, 
where we do not observe any KBO near $a\simeq50$-52 AU with near-circular and low-inclination 
orbit so far. In conclusion, based on the assumption of equal size objects and
Eq. (4), the remnant 
population predicted by the extra-planet model seems to contradict the present observations.

\section*{\bf 6. Conclusions}

\qquad
We explore the conventional Neptune migration model with one extra 
planet in the 3:2 MMR with Neptune during migration. After the
extra planet excites the primordial disk to a suitable state, we assume 
that this planet escapes from the disk in a short timescale and does 
not influence disk particles anymore. In all the our runs of the 
extra-planet model if we assume the primordial disk truncated at 
45 AU, the extra planet having the mass of $\sim1.0$ $M_\oplus$ 
and escaping roughly near the end of Neptune's migration produces the 
KBOs' orbital distribution which is most consistent with the observations.
The extra-planet model generates several 
observational features of the Kuiper belt that cannot be explained 
via the conventional Neptune migration model. These features are that 
(1) the high-inclination Plutinos with $i\simeq$ 
15{\textdegree}-35{\textdegree}; (2) the excitation of the classical 
KBOs, which have moderate eccentricities and inclinations; (3) 
the larger ratio of Neptune's 3:2 to 2:1 resonant particles, and 
the lower ratio of the Plutino to the classical KBO numbers, which may be more
consistent with observations; and (4) several 5:2 resonant 
particles. However, the simulation results in the discussion imply that 
this model is a low-probability event. 
In the extra-planet model, to understand the orbital evolution of the 
extra planet embedded in a massive disk, more realistic simulations 
and some analytic works are needed.
In addition to the low probability issue, two features produced by the
extra-planet model may be inconsistent with the observations. They are
the small number of low-inclination particles in the classical belt,
and the production of a remnant population with near-circular and
low-inclination orbit within $a\simeq50$-52 AU.
Although the extra-planet model can explain several features which
are unable to produce by the conventional Neptune migration model, 
it destroys the low-inclination population in the classical belt,
which is the easiest and a natural outcome of the conventional 
Neptune migration model. 
Since our simulations consist of particles with negligible mass, one
may wonder whether including the mass of particles could change this 
conclusion or not. Taking into account the mass of particles, 
we found that the dynamical friction suffered by the bodies of $\sim$ 
100-km size seems unlikely to generate more low-inclination 
particles in Gyr timescale, as indicated in Eq. (3). Whether the 
effect of physical collision between objects of different sizes 
mitigates this shortage has yet to explore.   
As for the remnant population, based on the assumption of equal size for all objects
and Eq. (4),
we may observe several tens of KBOs in this population according to the results
of the extra-planet model,
but to date no KBOs are observed to be with near-circular and 
low-inclination orbit near $a\simeq50$-52 AU.  

In summary, according to our present study, including one extra planet
in the conventional Neptune migration model as the scenario we explored 
here may be unsuitable because of the low probability, and the two drawbacks
mentioned above, although this model
can explain better several features which are hard to produce by the conventional 
Neptune migration model. The issues of the low probability and lack
of low-inclination classical KBOs are interesting and will be investigated further under a more
realistic consideration.

\section*{\bf Acknowledgments}

\qquad
We thank R. Gomes and two anonymous reviewers for a critical review 
and useful suggestions that greatly improved this paper.
We also would like to thank I.-G. Jiang for helpful discussion.
This work was supported by the National Science Council of the 
Republic of China under grant NSC 96-2628-M-007-012-MY3.

\section*{\bf References}

\begin{verse}
Brown, M.E., 2001. The inclination distribution of the Kuiper Belt.
Astron. J. 121, 2804-2814.\\
\end{verse}
\begin{verse}
Chambers, J.E., 1999. A hybrid symplectic integrator that permits 
close encounters between massive bodies. Mon. Not. R. Astron. Soc. 
304, 793-799.\\
\end{verse}
\begin{verse}
Chiang, E.I., Jordan, A.B., 2002. On the Plutinos and Twotinos of the 
Kuiper belt. Astron. J. 124, 3430-3444.\\
\end{verse}
\begin{verse}
Chiang, E.I., Jordan, A.B., Millis, R.L., Buie, M.W., Wasserman, 
L.H., Elliot, J.L., Kern, S.D., Trilling, D.E., Meech, K.J., Wagner, 
R.M., 2003. Resonance occupation in the Kuiper belt: Case examples of 
the 5:2 and Trojan resonances. Astron. J. 126, 430-443.\\
\end{verse}
\begin{verse}
Chiang, E., Lithwick, Y., Murray-Clay, R., Buie, M., Grundy, W., 
Holman, M., 2007. A brief history of trans-neptunian space. In: 
Reipurth, B., Jewitt, D., Keil, K. (Eds.), Protostars and Planets 
V Compendium. Univ. of Arizona Press, Tucson, pp. 895–911.
\end{verse}
\begin{verse}
de La Fuente Marcos, R. and de La Fuente Marcos, C., 2008. 
Confined chaotic motion in three-body resonances: trapping of 
trans-Neptunian material induced by gas-drag. Mon. Not. R. Astron. 
Soc. 388, 293-306.\\
\end{verse}
\begin{verse}
Fern{\'a}ndez, J.A., Ip, W.-H., 1996. Orbital expansion and resonant 
trapping during the late accretion stages of the outer planets. Planet. 
Space Sci. 44, 431-439.\\
\end{verse}
\begin{verse}
Gladman, B., Chan, C., 2006. Production of the extended scattered disk 
by rogue planets. Astrophys. J. 643, L135-L138.\\
\end{verse}
\begin{verse}
Gomes, R.S., 2009. On the origin of the Kuiper belt. Celest. Mech. Dyn. 
Astron., DOI: 10.1007/s10569-009-9186-5.\\
\end{verse}
\begin{verse}
Gomes, R.S., Morbidelli, A., Levison, H.F., 2004. Planetary migration 
in a planetesimal disk: why did Neptune stop at 30 AU? Icarus 170, 
492-507.\\
\end{verse}
\begin{verse}
Hahn, J.M., Malhotra, R., 1999. Orbital evolution of planets embedded 
in a planetesimal disk. Astron. J. 117, 3041-3053.\\
\end{verse}
\begin{verse}
Hahn, J.M., Malhotra, R., 2005. Neptune's migration into a stirred-up 
Kuiper belt: A detailed comparison of simulations to observations. 
Astron. J. 130, 2392-2414.\\
\end{verse}
\begin{verse}
Ida, S., Larwood, J., Burkert, A., 2000. Evidence for early stellar 
encounters in the orbital distribution of Edgeworth-Kuiper belt objects. 
Astron. J. 528, 351-356.\\
\end{verse}
\begin{verse}
Jiang, I.-G., Yeh, L.-C., 2004. The drag-induced resonant capture for 
Kuiper belt objects. Mon. Not. R. Astron. Soc. 355, L29-L32.\\
\end{verse}
\begin{verse}
Kenyon, S.J., 2002. Planet formation in the outer Solar System. Publ. 
Astron. Soc. Pacific 114, 265-283.\\
\end{verse}
\begin{verse}
Kenyon, S.J., Bromley, B.C., 2004. Stellar encounters as the origin of 
distant Solar System objects in highly eccentric orbits. Nature 432, 
598-602.\\
\end{verse}
\begin{verse}
Kenyon, S.J., Luu, J.X., 1999. Accretion in the early Kuiper belt. II. 
Fragmentation. Astron. J. 118, 1101-1119.\\
\end{verse}
\begin{verse}
Kenyon, S.J., Bromley, B.C., O'Brien, D.P., Davis, D.R., 2008. 
Formation and collisional evolution of Kuiper Belt objects. In: 
Barucci, M.A., Boehnhardt, H., Cruikshank, D.P., Morbidelli, A. 
(Eds.), The Solar System Beyond Neptune. Univ. Arizona Press, Tucson, 
pp. 293-313. 
\end{verse}
\begin{verse}
Levison, H.F., Stern, S.A., 1995. Possible origin and early dynamical 
evolution of the Pluto-Charon binary. Icarus 116, 315-339.\\
\end{verse}
\begin{verse}
Lykawka, P.S., Mukai, T., 2007a. Dynamical classification of 
trans-Neptunian objects: Probing their origin, evolution, and 
interrelation. Icarus 189, 213-232.\\
\end{verse}
\begin{verse}
Lykawka, P.S., Mukai, T., 2007b. Origin of scattered disk resonant 
TNOs: Evidence for an ancient excited Kuiper belt of 50 AU radius. 
Icarus 186, 331-341.\\
\end{verse}
\begin{verse}
Lykawka, P.S., Mukai, T., 2008. An outer planet beyond Pluto and 
the origin of the trans-Neptunian belt architecture. Astron. J. 
135, 1161-1200.\\
\end{verse}
\begin{verse}
Malhotra, R., 1995. The origin of Pluto's orbit: Implications for 
the solar system beyond Neptune. Astron. J. 110, 420-429.\\
\end{verse}
\begin{verse}
Murray, N., Holman, M., Potter, M., 1998. On the origin of chaos in 
the asteroid belt. Astron. J. 116, 2583-2589.\\
\end{verse}
\begin{verse}
Nesvorn{\'y}, D., Morbidelli, A., 1998a.
An analytic model of three-body mean motion resonances.
Celest. Mech. Dynam. Astron. 71, 243-271.\\
\end{verse}
\begin{verse}
Nesvorn{\'y}, D., Morbidelli, A., 1998b.
Three-body mean motion resonances and the chaotic structure of the 
asteroid belt. Astron. J. 116, 3029-3037.\\
\end{verse}
\begin{verse}
Nesvorn{\'y}, D., Roig, F., 2000. Mean motion resonances in the 
trans-Neptunian region I. The 2:3 resonance with Neptune. Icarus 
148, 282-300.\\
\end{verse}
\begin{verse}
Petit, J.-M., Morbidelli, A., Valsecchi, G.B., 1999. Large scattered 
planetesimals and the excitation of the small body belts. Icarus 141, 
367-387.\\
\end{verse}
\begin{verse}
Stern, S.A., 1991. On the number of planets in the outer solar system - 
Evidence of a substantial population of 1000-km bodies. Icarus 90, 
271-281.\\
\end{verse}
\begin{verse}
Trujillo, C.A., Brown, M.E., 2001a. The radial distribution of the Kuiper 
belt. Astrophys. J. 554, L95-L98.\\
\end{verse}
\begin{verse}
Trujillo, C.A., Jewitt, D.C., 2001b. Properties of the trans-Neptunian 
blet: Statistics from the Canada-France-Hawaii telescope survey.
Astron. J. 122, 457-473.\\
\end{verse}
\begin{verse}
Zhou, L.-Y., Sun, Y.-S., Zhou, J.-L., Zheng, J.-Q., Valtonen, M., 2002. 
Stochastic effects in the planet migration and orbital distribution of 
the Kuiper belt. Mon. Not. R. Astron. Soc. 336, 520-526.\\
\end{verse}

\clearpage

\begin{figure}
\begin{center}
\includegraphics[width=180mm]{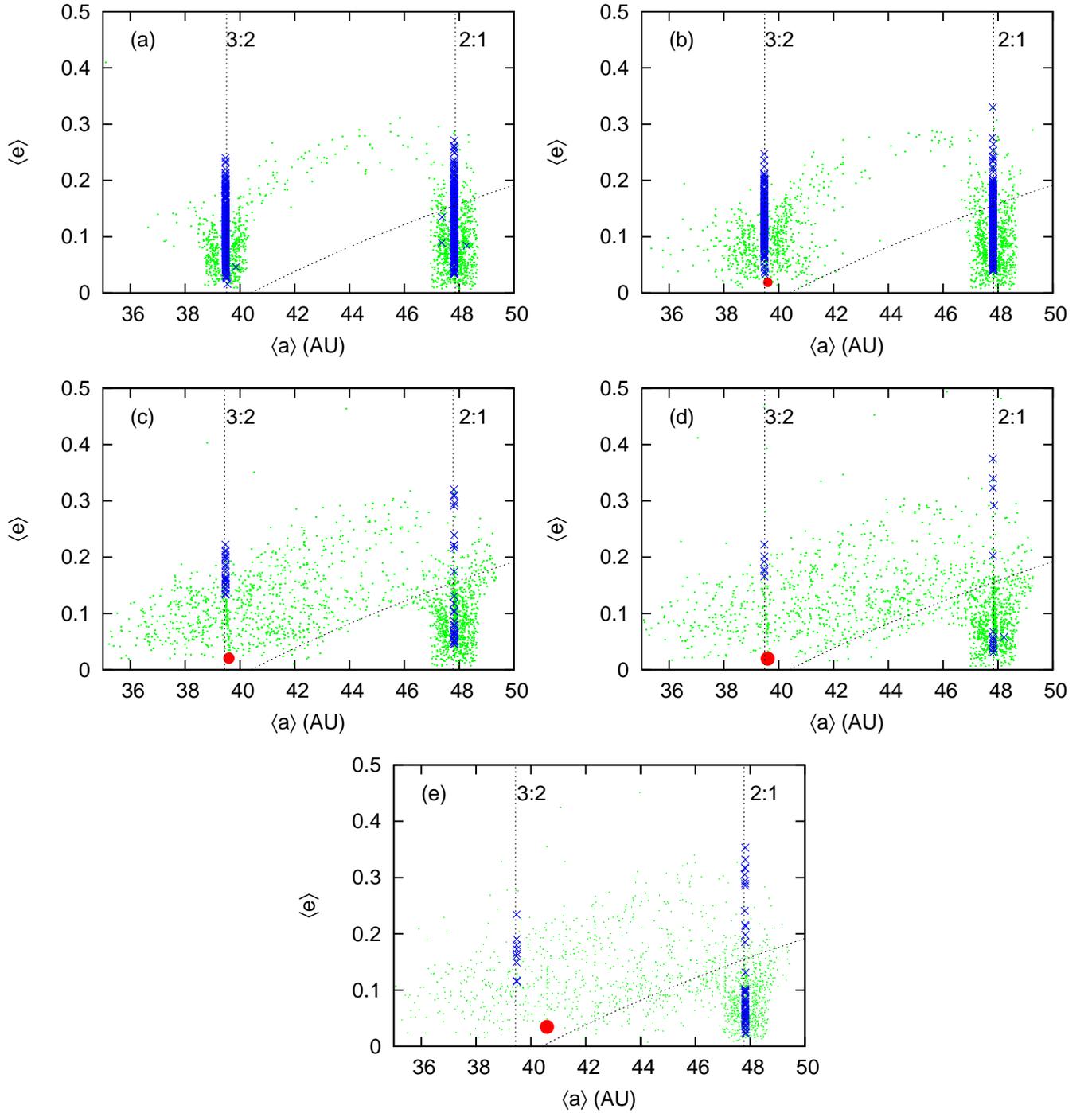}
\caption
{Time-averaged semimajor axis $\langle a\rangle$ and eccentricity 
$\langle e\rangle$ at the end of 8.25 Myr simulation over a time 
interval of 0.825 Myr. (b), (c) and (d) are the runs including one 
extra planet in Neptune's 3:2 MMR with the mass of 0.1, 0.5 and 1.0 
$M_\oplus$, respectively. (a) is the run without extra planet for 
comparison. Non-resonant particles are represented by green dots. 
Neptune's 3:2 and 2:1 resonant particles are marked by blue crosses. 
Red filled circle marks the extra planet. The perihelion of 40.4 AU is 
plotted by dashed line. Two vertical dashed lines represent the 
locations of the exterior 3:2 and 2:1 MMR with Neptune. In figure (e), 
one extra planet with 1.0 $M_{\oplus}$ is set one AU farther beyond 
the exact Neptune's 3:2 MMR at the beginning.}
\end{center}
\end{figure}

\begin{figure}
\begin{center}
\includegraphics[width=120mm]{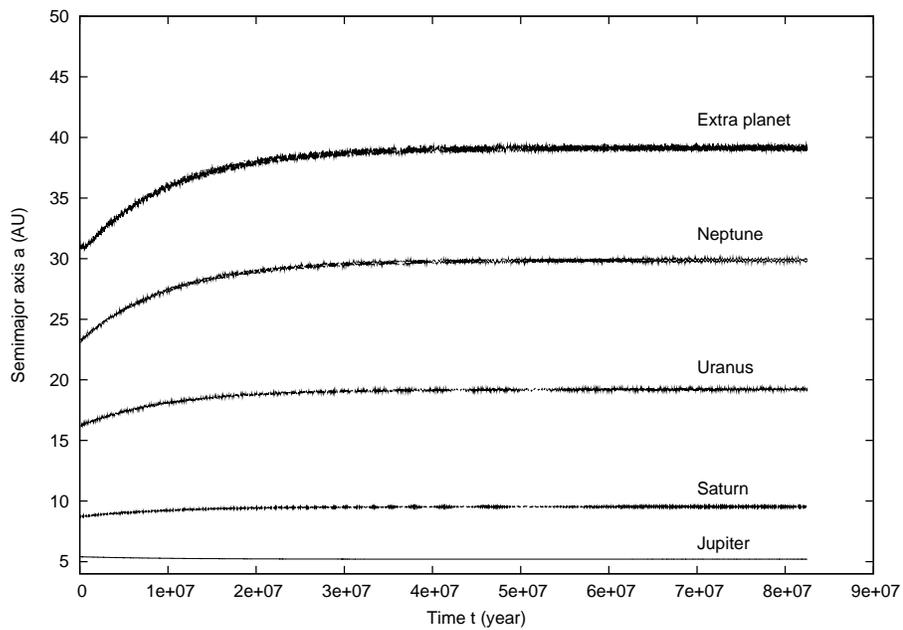}
\caption
{Semimajor axes variation with time during the planet migration for 
the four Jovian planets and the extra-planet A with the mass of 1.0 $M_\oplus$ 
as an example. The smooth migration of the four Jovian planet are 
driven by the artificial force. The extra planet was captured in
Neptune's 3:2 MMR and moved outward together with Neptune.}
\end{center}
\end{figure}

\begin{figure}
\begin{center}
\includegraphics[width=180mm]{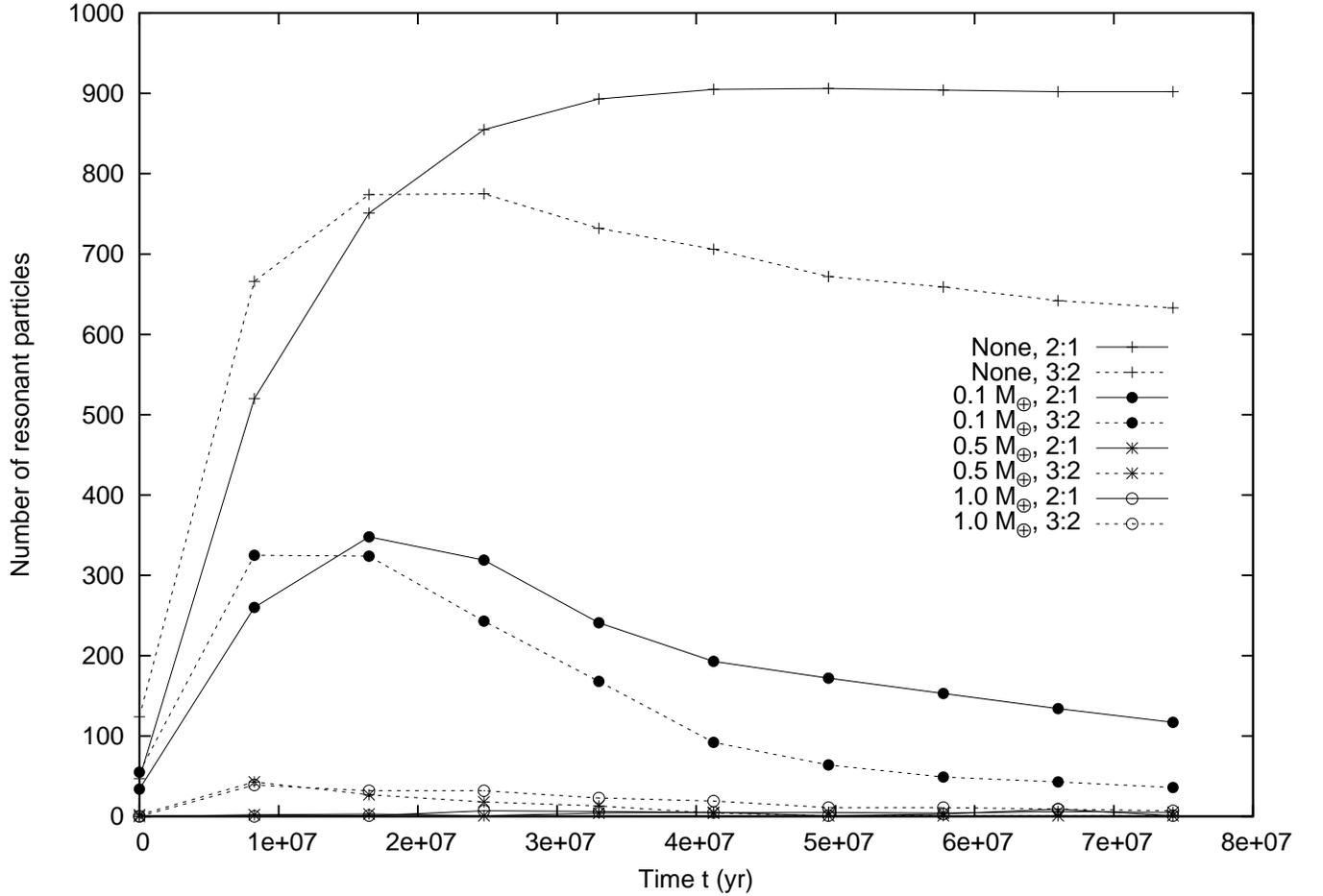}
\caption
{Neptune's 3:2 and 2:1 resonant particle numbers changing with time 
for the particle-set a. The three runs with different masses extra planet 
and the one run without extra planet are shown. The time interval for 
resonant particle identification is 8.25 Myr. Dashed and solid lines 
indicate the 3:2 and 2:1 resonant particle numbers, respectively. 
The runs with the extra planet of 0.1 $M_\oplus$, 0.5 $M_\oplus$ and 
1.0 $M_\oplus$ are indicated with filled circles, stars and open 
circles, respectively. The run without one additional planet is 
shown by crosses. In the runs containing 0.5-$M_\oplus$ and 
1.0-$M_\oplus$ extra planets, the 2:1 resonant particle numbers are 
always less than ten.}
\end{center}
\end{figure}

\begin{figure}
\begin{center}
\includegraphics[width=180mm]{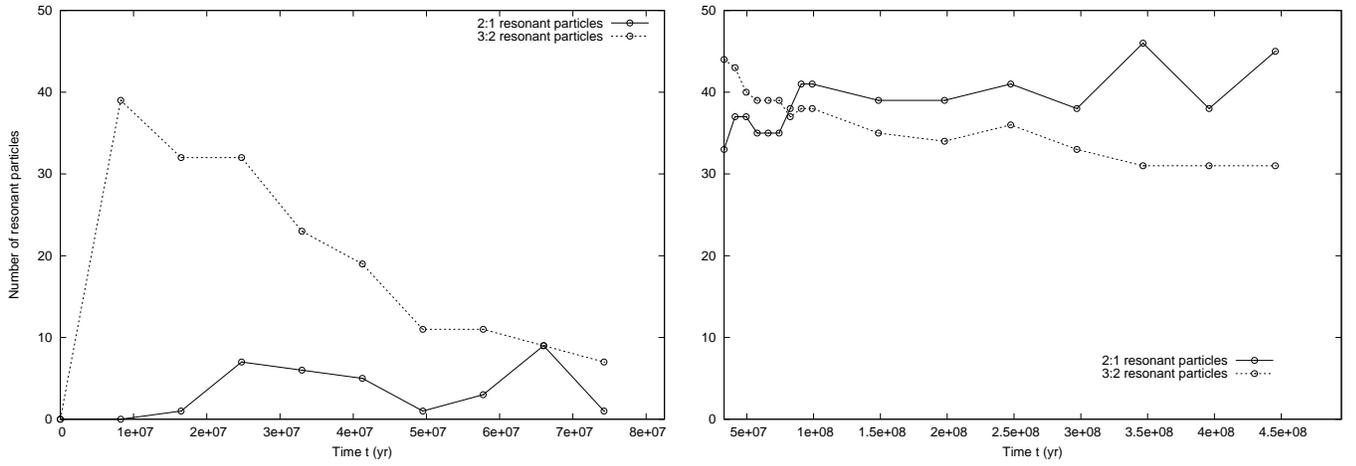}
\caption
{Neptune's 3:2 and 2:1 resonant particle numbers varying with time. 
The left panel is just an enlarged version of part of 
Fig. 3 for the run containing 1.0-$M_\oplus$ extra planet. We reproduce 
the same run but removed this additional planet at 33 Myr, and then 
continued integration to 495 Myr. Right panel plots this long-term
integration and starts from t = 33 Myr. Dashed and solid lines 
represent Neptune's 3:2 and 2:1 resonant particle numbers, 
respectively.}
\end{center}
\end{figure}

\begin{figure}
\begin{center}
\includegraphics[width=180mm]{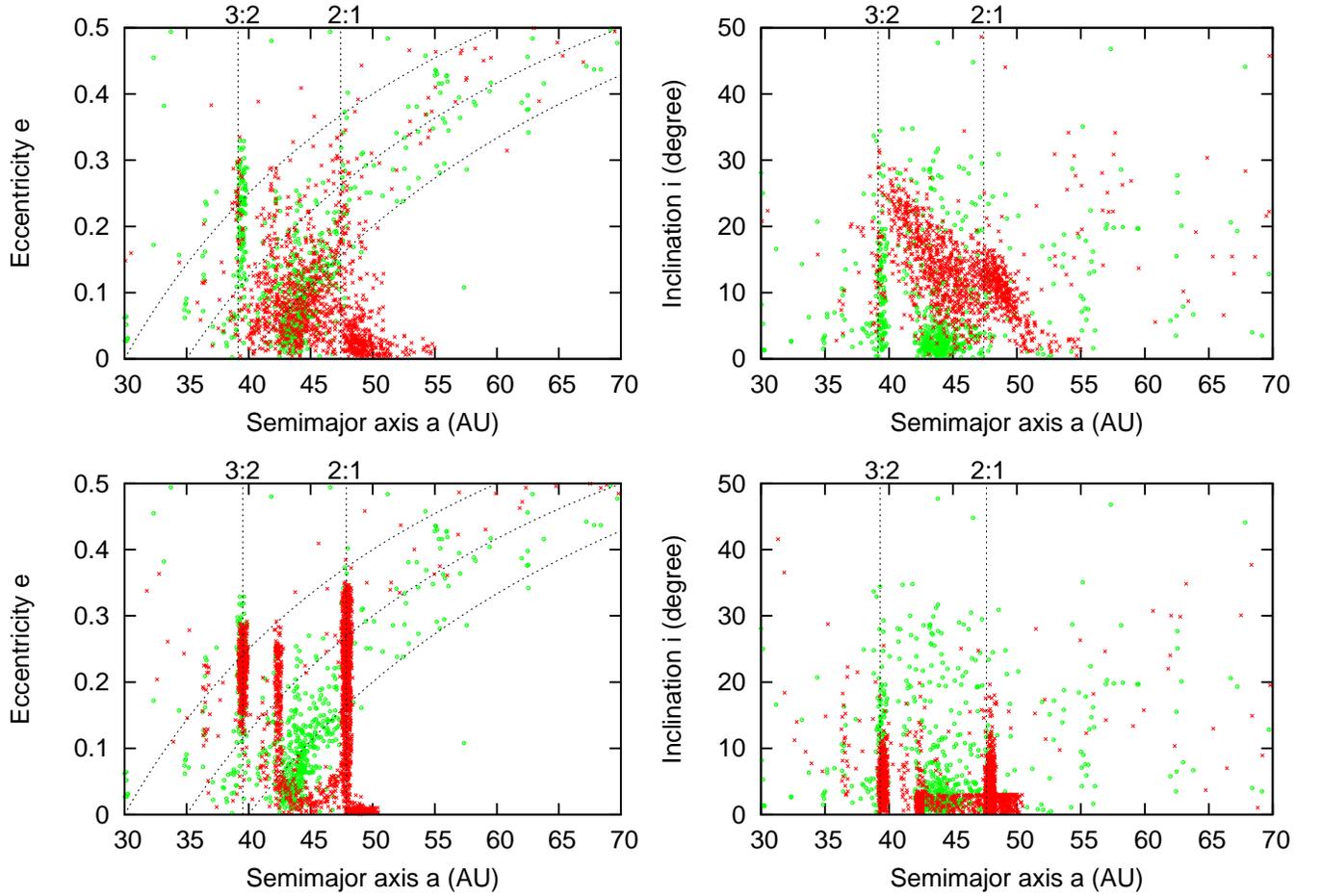}
\caption
{The orbital elements for the runs containing the particle-set a and 
the 1.0-$M_\oplus$ extra-planet A. This figure plots the orbital 
distribution of test particles (red crosses) at the end of 495 Myr for 
the extra-planet model (top panels) and the conventional migration 
model (bottom panels), respectively. The extra-planet A was removed at 
33.0 Myr in the extra-planet model. Neptune's 3:2 and 2:1 MMR are 
indicated with vertical lines. The perihelia of 30, 35 and 40 AU are 
shown with curved dashed lines. The observed KBOs with multi-opposition 
are also plotted with green circles (from the Minor Planet Center 
database of July, 10, 2007).}
\end{center}
\end{figure}

\begin{figure}
\begin{center}
\includegraphics[width=180mm]{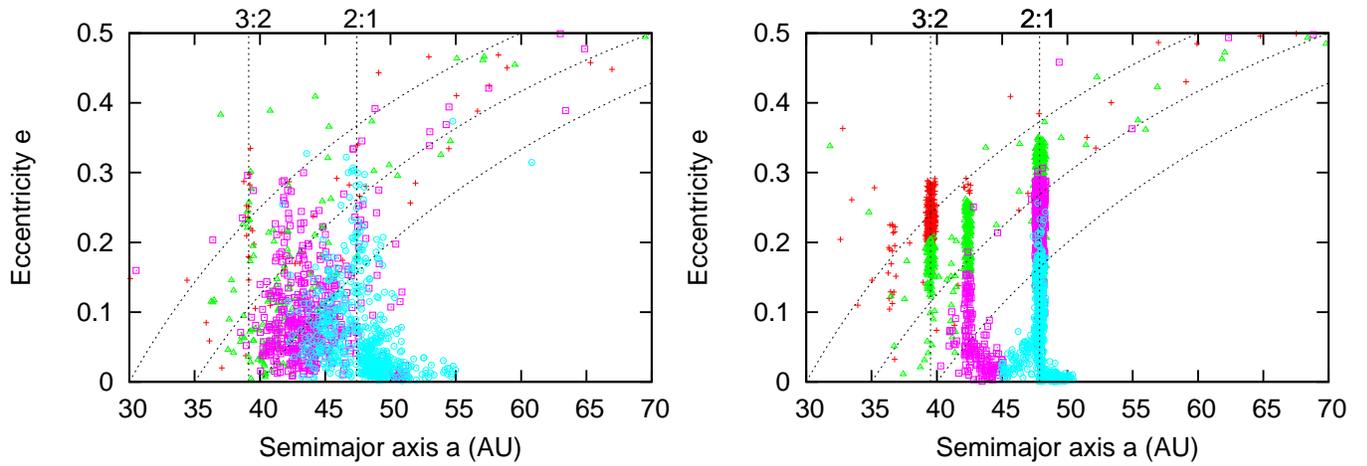}
\caption
{This figure plotting the orbital elements which are the same as the left
two panels of Fig. 5, but we label particles with their initial 
semimajor axes. Crosses (red), triangles (green), squares (purple) 
and circles (blue) indicate the particles with initial semimajor 
axes from 24-35 AU, 35-40 AU, 40-45 AU and 45-50 AU, respectively.}
\end{center}
\end{figure}

\begin{figure}
\begin{center}
\includegraphics[width=180mm]{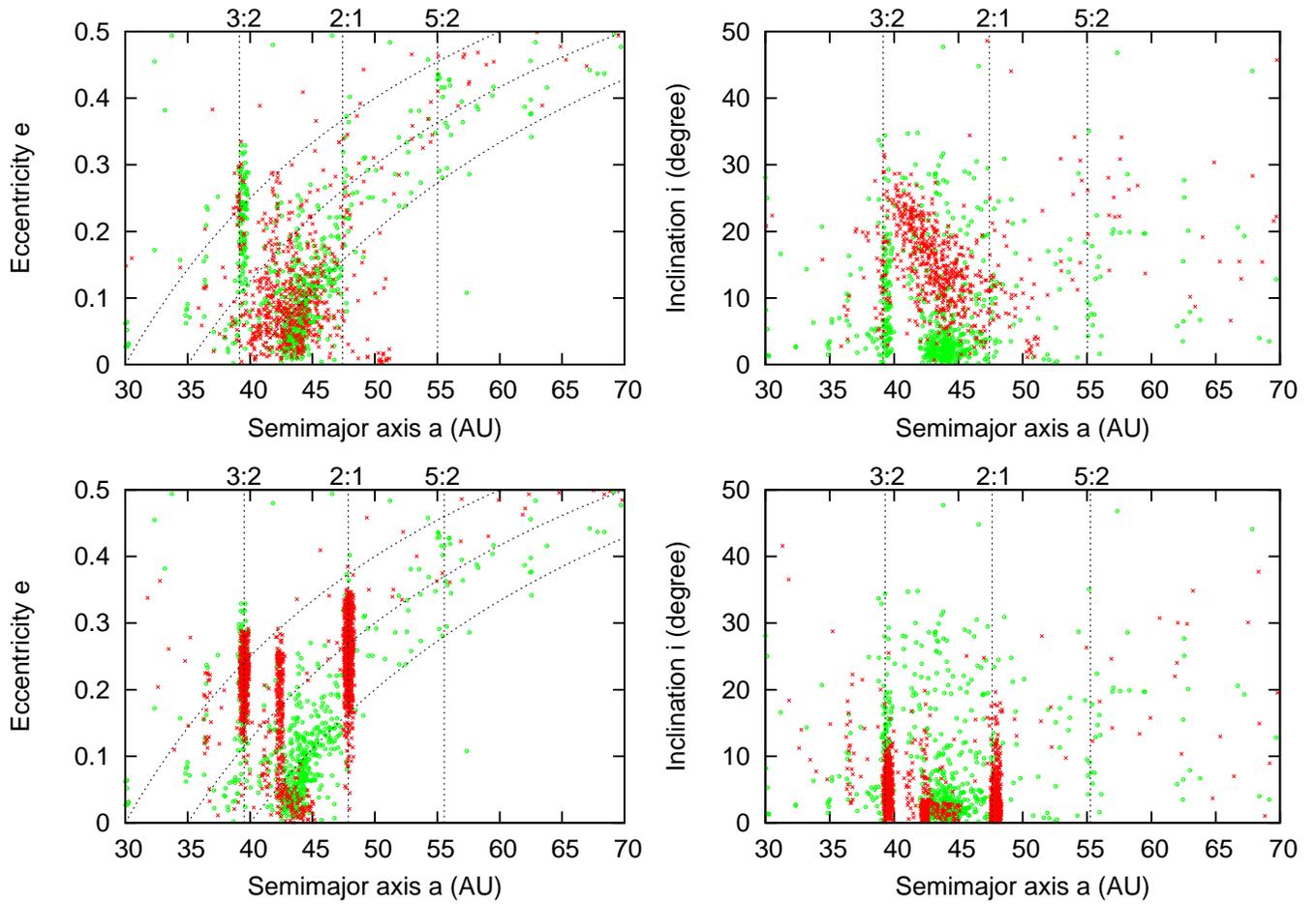}
\caption
{The same figure as Fig. 5 but with initial disk truncated at 
$a=45$ AU. The 5:2 MMR with Neptune is also labeled with vertical
dashed line.}
\end{center}
\end{figure}

\begin{figure}
\begin{center}
\includegraphics[width=120mm]{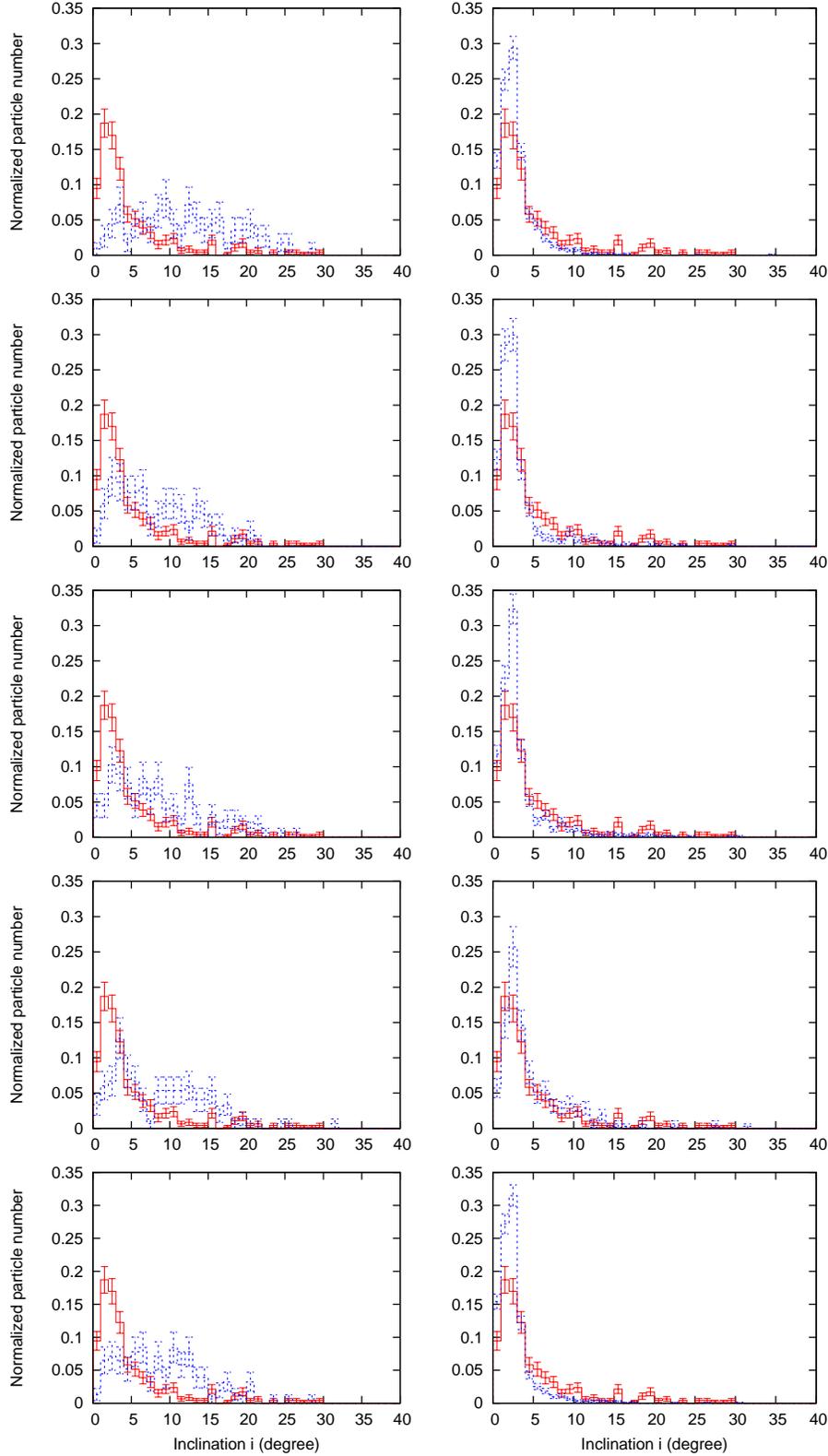}
\caption
{This figure showing the ecliptic inclination distribution for the 
results of the extra-planet model (left panels) and those of the 
conventional migration model (right panels) at the simulation end of 
495 Myr. Top to bottom figures are the runs containing the 
particle-set a, b, c, d and e, respectively. The extra planets with 
1.0 $M_\oplus$ were removed at 33 Myr in the extra-planet model. We 
chose particles with ecliptic latitudes $\beta\le3.0${\textdegree} 
and perihelion $q\le45.0$ AU. Vertical axes represent
particle number normalized to total particle number under above 
constraints. Simulated results and observed KBOs with multi-opposition 
are plotted by blue dashed and red solid lines, respectively. The 
Poisson standard deviations are also shown.}
\end{center}
\end{figure}

\begin{figure}
\begin{center}
\includegraphics[width=150mm]{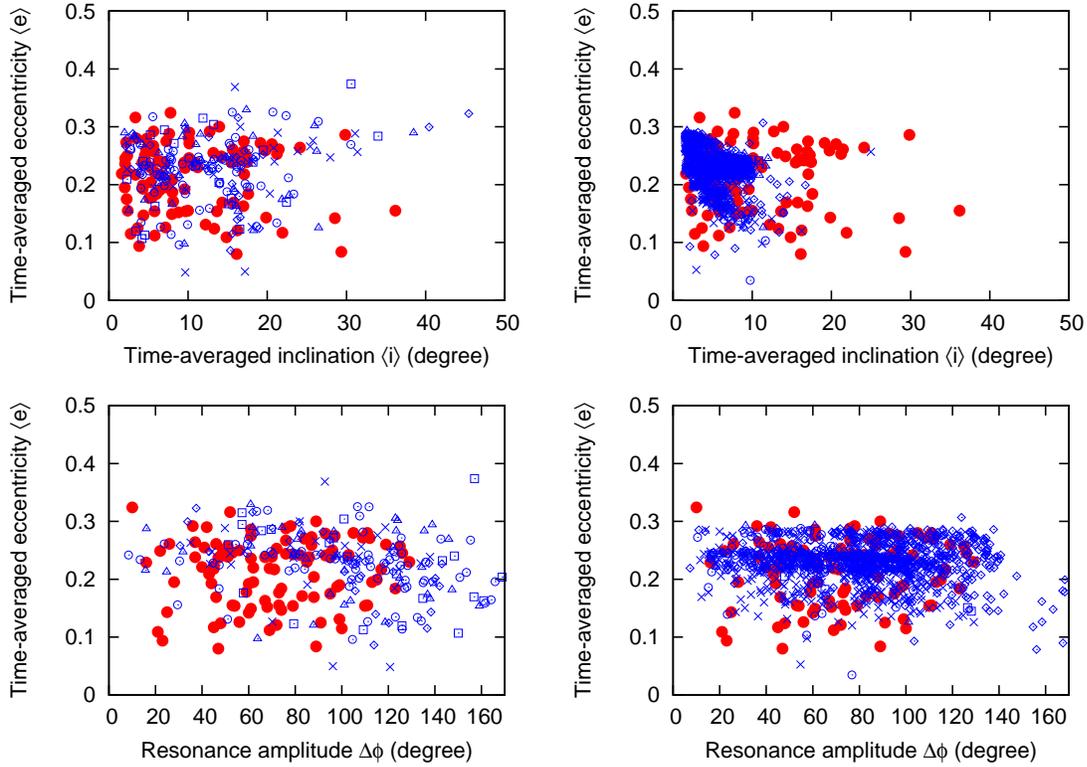}
\caption
{This figure showing time-averaged orbital elements 
$\langle e\rangle$, $\langle i\rangle$ and Neptune's 3:2 
resonant amplitude $\Delta\phi$ for simulated Plutinos 
(bule symbols) at the end of 495 Myr simulation, where the runs 
containing particle-set a, b, c, d and e are labeled by crosses, 
squares, open circles, triangles and diamonds, respectively. Left two 
panels are the results of the extra-planet models, where the 
1.0-$M_\oplus$ extra planet was removed at 33 Myr. Right two panels 
are the results of the conventional migration model. 100 observed 
Plutinos are also plotted with red filled circles from the 
identification of Lykawka and Mukai (2007a).}
\end{center}
\end{figure}

\begin{figure}
\begin{center}
\includegraphics[width=150mm]{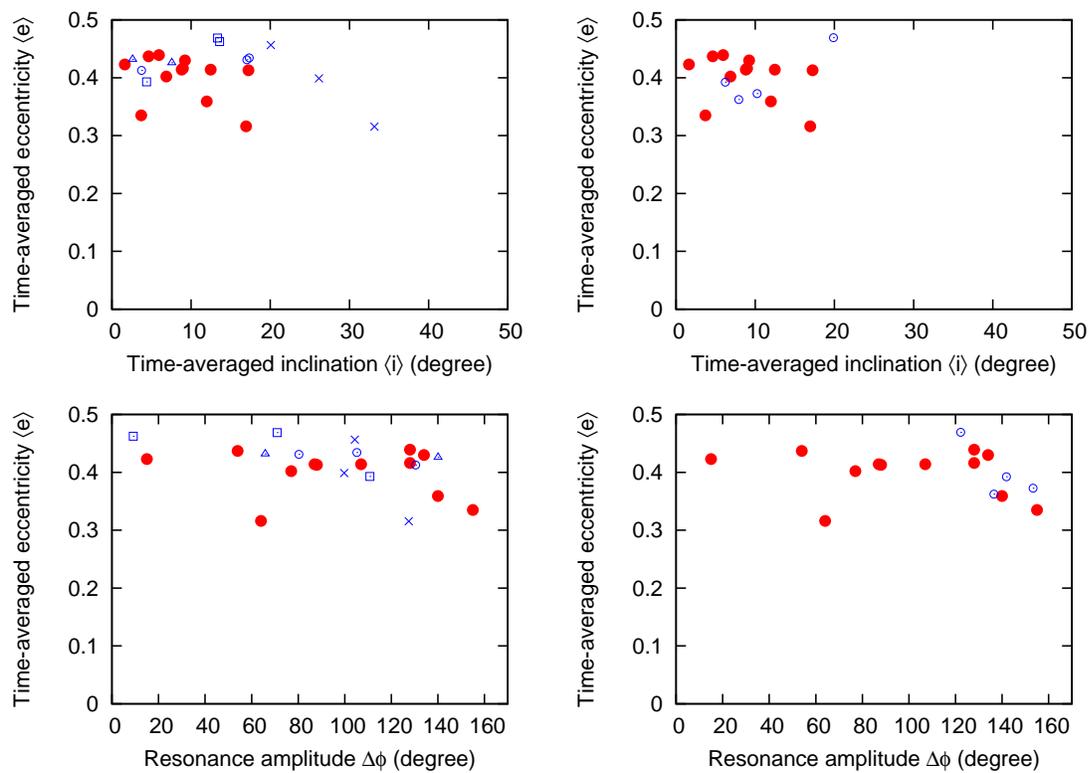}
\caption
{The same with allocation of Fig. 9 but for Neptune's 5:2 resonant 
particles.}
\end{center}
\end{figure}

\begin{figure}
\begin{center}
\includegraphics[width=120mm]{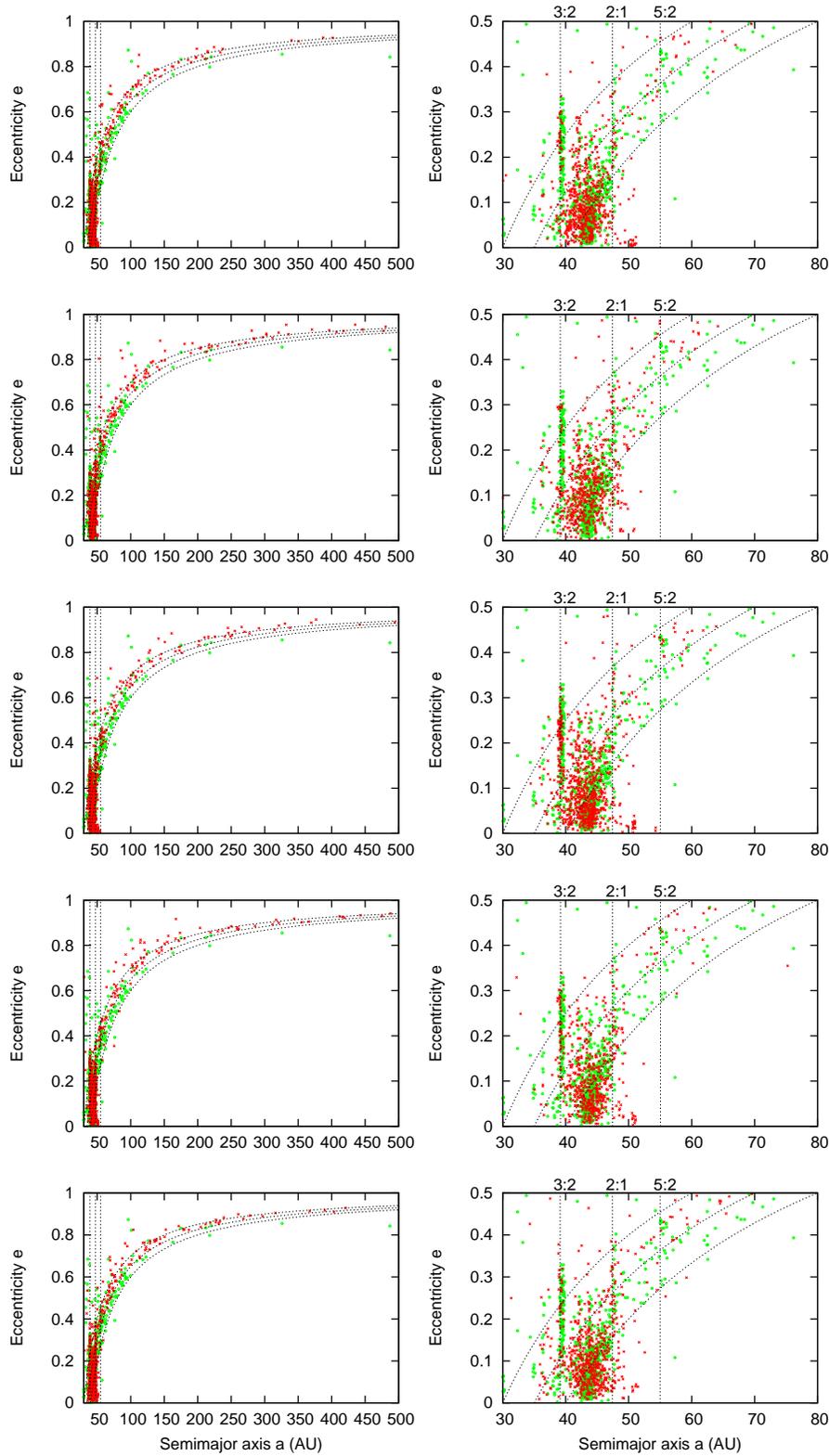}
\caption
{The $a$-$e$ distribution for the extra-planet model with one 
1.0-$M_\oplus$ extra planet removing at 33 Myr. From top to bottom 
are the runs containing particle-set a, b, c, d and e, respectively. 
The left and the right panels are the same figures but with different 
scale for semimajor axis. Simulated particles and observed KBOs are 
shown with red crosses and green circles, respectively. Three vertical 
lines indicate Neptune's 3:2, 2:1 and 5:2 MMR. Perihelia of 30, 35 
and 40 AU are represented by dashed curves.}
\end{center}
\end{figure}

\begin{figure}
\begin{center}
\includegraphics[width=180mm]{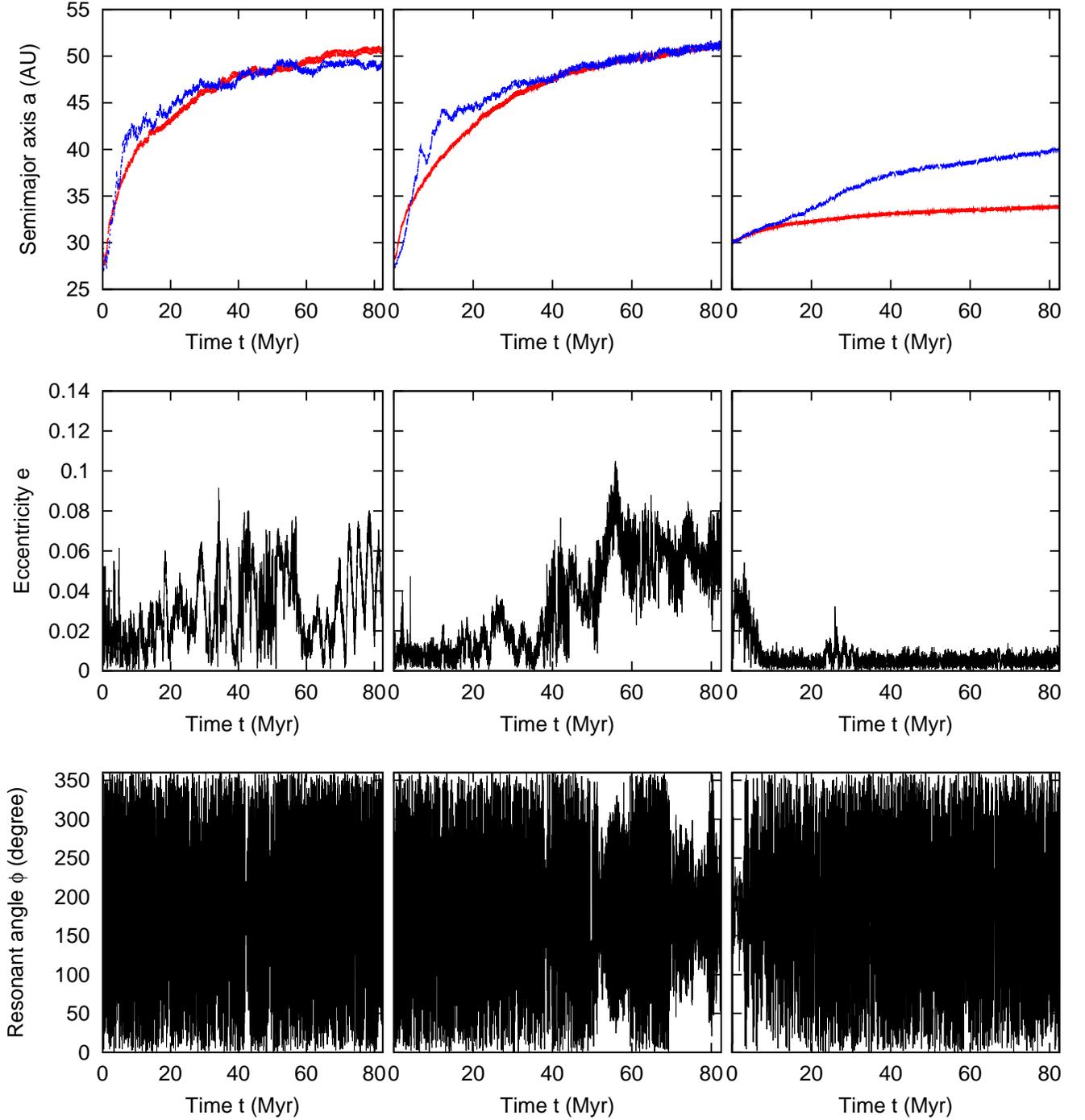}
\caption
{Semimajor axes $a$, the eccentricities $e$ and Neptune's 3:2 resonant
angles $\phi$ varying with time for the three massive-disk runs. Left panels show 
one run with 4000 massive particles and disk mass having 80 $M_\oplus$
as initial conditions. Middle panels are one run consisting of 
16000 massive particles and the disk mass of 80 $M_\oplus$. A low-mass-disk
run with 15 $M_\oplus$ and 16000 massive particles is shown in 
right panels. $t$-$a$ diagrams plot semimajor axis of the extra 
planet (blue curve) and that of Neptune's 3:2 MMR (red curve) changing with time. 
The capture of the extra planet by the 3:2 MMR can be identified by librating 
resonant angles.}
\end{center}
\end{figure}

\end{document}